\documentclass[twocolumn]{article}
\usepackage{graphicx,amsmath,amssymb,amsfonts,amsthm,authblk}

\usepackage{endnotes}
\usepackage{setspace}
\usepackage{verbatim}
\usepackage[left=1in, right=0.7in, bottom=0.9in, top=0.5in]{geometry}
\usepackage{times}
\usepackage{helvet}
\usepackage[switch,columnwise]{lineno}
\usepackage{courier}
\usepackage{bm}
\usepackage{url}
\usepackage{dcolumn}
\usepackage{multirow}

\date{}

\begin{document}
\title{Impact of commuting on disease persistence in heterogeneous metapopulations}
\author[1,2]{G.~Rozhnova}
\author[1]{A.~Nunes}
\author[2]{A.~J.~McKane}

\affil[1]{Centro de F{\'\i}sica da Mat{\'e}ria Condensada and Departamento de F{\'\i}sica, Faculdade de Ci{\^e}ncias da Universidade de Lisboa, P-1649-003 Lisboa Codex, Portugal}
\affil[2]{Theoretical Physics Division, School of Physics and Astronomy, University of Manchester, Manchester M13 9PL, United Kingdom}

\maketitle

\begin{abstract}
We use a stochastic metapopulation model to study the combined effects of seasonality and spatial heterogeneity on disease persistence. We find a pronounced effect of enhanced persistence associated with strong heterogeneity, intermediate coupling strength and moderate seasonal forcing. Analytic calculations show that this effect is not related with the phase lag between epidemic bursts in different patches, but rather with the linear stability properties of the attractor that describes the steady state of the system in the large population limit.
\end{abstract}

\section{Introduction}

Demographic stochasticity due to the probabilistic nature of events such as births, deaths, mating and disease transmission, plays a major role in the dynamics of small populations. Its impact was acknowledged more than fifty years ago by Bartlett \cite{bartlett:57}, who introduced the concept of critical community size. Originally defined as the population size above which the expected time to fade out after an epidemic exceeds a certain period, it is usually taken in more general terms as the threshold population for (a given definition of) disease persistence. It became a central concept in epidemiology, much revisited in several attempts to provide less arbitrary definitions and to reconcile theoretical estimates with data \cite{keelinggrenfell:02,Nasell:2004}. Threshold levels of host abundance are equally important in ecology, a context in which the idea of the stochastic Allee effect was introduced \cite{Lande:1998} to represent demographic stochasticity.

The fact that many natural populations experience annual abundance troughs establishes an obvious connection between average population size and extinction probability, on one hand, and seasonality, on the other \cite{Yorke:1979}. Indeed, the annual and multiannual incidence patterns of many infectious diseases show that seasonality is a key ingredient in the overall dynamics of these diseases. Despite the mathematical difficulties involved, theoretical studies have therefore tried to take seasonality into account ever since the earliest efforts \cite{Soper:1929}. The complex interplay between seasonal forcing and the system's nonlinearities is nowadays reasonably well understood, setting the stage for the additional layer of complexity that arises from demographic stochasticity \cite {stone:2007,conlanBirth:2007,Mantilla:2010}.

Another key ingredient for population persistence is spatial structure and heterogeneity. Spatial structure was first addressed using reaction-diffusion equations that successfully modelled the spread of the epizootic in animal borne diseases  \cite{Kallen:1985}. In these models, inspired by physical systems, the interactions are local and the population is distributed on a plane. More recently, developments that explore the role of individual mobility and long range interactions have come up in the form of metapopulation models, where a number of typically weakly interacting units represent well mixed homogeneous population patches \cite{harwood:1997,Hanski:1998,Levins:1969,Lloyd:1996,Riley:2007,Balcan:2009}. A long standing idea associated with the concept of metapopulation is that persistence is favoured in a fragmented population, provided that movement between patches accompanies spatial dispersion \cite{Bolker:1995,Hanski:1999}. This idea has recently been shown to be less straightforward than previously thought \cite{Hagenaars:2004, Jesse:2011}.

Among many aspects treated in these studies on spatially extended systems, the degree of synchrony of population abundance oscillations has received special attention as it has been considered the main determinant of persistence \cite{Kleczkowski:1995,Blasius:1999,grassly:06}. With few exceptions associated with chaotic oscillations, it has been found that a small amount of coupling between population patches is enough to induce synchrony \cite{earnrohani:1998,Viboud:2006,grenfell:01,Lloyd:1996}. Although they are usually called spatially heterogeneous, the metapopulation models in these studies assume the same, or very similar, parameter values for the different population patches, and we will refer to them as 'uniform', keeping the term 'heterogeneous' for extended systems that include significant parameter variation across different patches. In line with available data for large urban populations \cite{Viboud:2006,Mantilla:2010}, the synchronized oscillations in uniform systems are moreover found to be in phase between patches, or, in the case of the 2-year cycle typical of measles, in phase opposition. This is in contrast with the results for heterogeneous systems, where synchronous states may correspond to intermediate phase lags \cite{Blasius:1999,rozhnova:12b}.

In this paper we present an extensive computational study of the combined influence of seasonality and heterogeneity on disease persistence. The basic unit of our model, which we will call a 'city', is formed by a number of individuals undergoing well-mixed stochastic infection dynamics whose parameters are specific to that city and may 
present seasonal variation. The number of individuals that interact in this way may comprise commuters from another city, as well as the residents of that city. Disease persistence is measured over sets of stochastic simulations of the model. We find that it depends in a nonintuitive way both on the level of seasonality and on the magnitude of the flow of commuters, with a pronounced enhancement of persistence induced by strong heterogeneity at intermediate coupling strengths. We also find that the epidemic phase lags generated by city heterogeneity have no significant effect on disease persistence. 

For the unforced case, an analytic description of the incidence fluctuations based on van Kampen's expansion was shown to give good quantitative results for moderate system sizes \cite{rozhnova:11,rozhnova:12b}. Using this approximation, summarized in the Supplementary Material (SM), it can be seen that this increase in persistence is instead related with the stability properties of the attractor that describes the steady state of the system in the large population limit.

\section{Methods}

\subsection{Model}
\label{sec:model}

In this section, we briefly present the metapopulation susceptible-infectious-recovered (SIR) stochastic model introduced in \cite{rozhnova:11,rozhnova:12b} to describe several interacting cities, which are population patches where interactions between individuals are taken to be well mixed.

The SIR model consists of three classes of individuals: susceptibles, infected and recovered. We denote their number among the residents of city $k$ by $S_k$, $I_k$, $R_k$, respectively. These numbers change due to birth, death, infection and recovery, which in the stochastic version of the model are taken as stochastic events with certain rates. As usual when working with time scales for which there are no major demographic changes, we assume that the number of individuals that reside in city $k$, $N_k$, is fixed, so that $S_k$ and $I_k$ together completely determine the state of city $k$. The birth/death rate $\mu$ is taken to be constant, and infected individuals recover also at a constant rate $\gamma$. When a given disease spreads in a city, the rate of infection is proportional to the number of encounters between susceptibles and infected that take place in that city, which in turn, assuming that in the city the population is well mixed, is proportional to the product of the number of susceptibles and the number of infected in that city. Now these numbers should take into account the flow of commuters from and to that city. In the simplest version of the model, we will assume that the coupling between cities 1 and 2 may be described by a single parameter, $f$, which is the fraction of the number of residents of each class of city 1 (respectively, 2) that are present in city 2 (respectively, 1) at any given time. The parameter $f$ must be interpreted as the overall fraction of time that an individual from one city spends in the other city, averaged over all types of stays with their typical frequencies and durations. In general, $f$ should be taken class and city dependent (see SM), but we will explore here only the simplest case. 

The usual SIR rate of infection then becomes, for susceptible residents of city 1 while in city 1,
$$
\beta_1 (1-f)S_1 \left[(1-f)I_1 + f I_2\right]/M_1,
$$ 
where $\beta_1$ is a parameter that reflects the urban characteristics of city 1 through the rate of encounters they elicit, and $M_1 = (1-f)N_1 + f N_2$ is the number of individuals present in city 1 at any given time. The rate of infection of susceptible residents of city 1 while in city 2 will be given by
$$
\beta_2 f S_1 \left[(1-f)I_2 + f I_1\right]/M_2,
$$ 
with $M_2 = (1-f)N_2 + f N_1$. Similar expressions hold for the rates of infections taking place in city 2.

Our mechanistic model thus leads to represent the interaction between population patches as a weighted distribution of their respective forces of infection. Along with other metapopulation models based on a description of the underlying mobility patterns \cite{keeling:02,House:2010}, it extends the traditional phenomenological modelling of interacting population patches by means of a single coupling parameter \cite{Lloyd:1996}, with the important difference that the parameters $\beta_k$ are allowed to differ from patch to patch, so that spatial heterogeneity does not come from 'patchiness' of the population only.

The parameter $\beta_k$ may be time dependent to represent seasonal variability of social intercourse, or of other ingredients such as for instance weather conditions that influence the rate of infectious contacts. We will consider a time dependence of the form
$
\beta_k(t) = \beta_{k}^{0} (1 + \epsilon \cos 2 \pi t),
$
where $t$ is the time measured in years and $\epsilon$ represents the amplitude of seasonal forcing. More realistic forcing terms that include a representation of school term calendars are commonly found in the literature on childhood infectious diseases (e.g. \cite{keelinggrenfell:02}), but we expect the overall picture revealed by varying $\beta_{k}^{0}$ and $\epsilon$ to be largely independent of the particular form of the periodic forcing.

With these assumptions, the stochastic process is governed by the master equation for the time evolution of $P_{{\bf n}}(t)$, the probability distribution for finding the system in state ${\bf n}$ at time $t$ \cite{kampen:1981}:
\begin{equation}
\frac{d P_{{\bf n}}(t)}{dt} = \sum_{{\bf n}' \neq {\bf n}} \sum_{\alpha }
\left[ T_{\alpha }({\bf n} | {\bf n}')P_{{\bf n}'}(t) - T_{\alpha }({\bf n}' | {\bf n}) P_{{\bf n}}(t) \right],
\label{master}
\end{equation}
where ${\bf n}$ denotes the state of the system given by the numbers of infected and susceptibles in each city and
$T_{\alpha }({\bf n} | {\bf n}')$, are the (possibly time dependent) transition rates from the state ${\bf n}'$ to the state ${\bf n}$  that result from the birth-death, recovery and infection processes. These rates are given explicitly in the SM.

\subsection{Theoretical Analysis and Simulations}

A deterministic description of the model leads to a set of ordinary differential equations (ODEs) for the evolution of the fractions $s_k=S_k/N_k$, $i_k=I_k/N_k$ of susceptible and infected individuals in each city. The behaviour of the stochastic system approaches this description in the limit of infinite population sizes where fluctuations can be neglected. Otherwise, for large but finite systems $s_k$ and $i_k$ will fluctuate around the solutions of the deterministic ODEs. Using van Kampen's system size expansion \cite{kampen:1981}, these fluctuations are approximately described by Langevin equations, which for two cities can be written in compact form as
\begin{eqnarray}
\dfrac{d z_J}{dt}&=& \sum _{K=1}^{4} {\cal J}_{JK} z_K + \eta_J (t), \; J=1, ...,4,
\label{langevin} 
\end{eqnarray}
where ${\bf z} = (z_1, z_2, z_3, z_4)= (x_1, x_2, y_1, y_2)$, $x_k$ and $y_k $ are the relative fluctuations of the number of susceptible and infected in city $k$, and $\eta_J (t)$ are zero mean Gaussian noise terms. The derivation  of these analytic approximations, as well as explicit expressions for its coefficients and for the noise correlation function, are given in the SM.

Neither the full stochastic system, Eq. (\ref{master}), nor the deterministic ODEs can be solved analytically. The latter are however amenable to qualitative analysis, which we will use, together with numerical integration, to study their attractors and their stability. As to the former,  Eq. (\ref{langevin}) can be used to compute the state variables fluctuation power spectrum and phase spectrum \cite{rozhnova:12b}, which determine the amplification (the overall power of the fluctuation spectrum), the coherence (the fraction of the total power in a small frequency range around the dominant frequency), and the phase lag between cities. Details are given in the SM.

The simulations implement the stochastic process described in the preceding section on the state variables $(S_k,I_k)$, using Gillespie's algorithm adapted to account for time dependent reaction rates \cite{gillespie:76}. We compute the average extinction time (AET) for a metapopulation as the average over $10^3$ simulations of the time it takes for the system to reach a state with no infectives, starting from different initial conditions near the equilibrium of the deterministic counterpart of the model. The simulation time was taken long enough for all runs to eventually become extinct.

The crucial epidemiological parameter $R_0$ of the unforced single city deterministic SIR model, the basic reproductive ratio, is given by $R_0=\beta/(\gamma + \mu)$, where $\beta $ is defined as usual through the infection rate $\beta S I$ in that isolated city. It is well known that $R_0=1$ is the critical value that separates the trivial regime, where the disease dies out, from the endemic regime, where the disease persists for ever. We will take $R_0$, $\epsilon$ and $f$ as the free parameters in this study. The parameters $\mu$ and $\gamma$ are kept fixed at $1/50$ yr and $1/13$ d.

\section{Results}

\begin{figure*}
\centering
\includegraphics[trim=0cm 0cm 0cm 0cm, clip=true, width=0.361\textwidth]{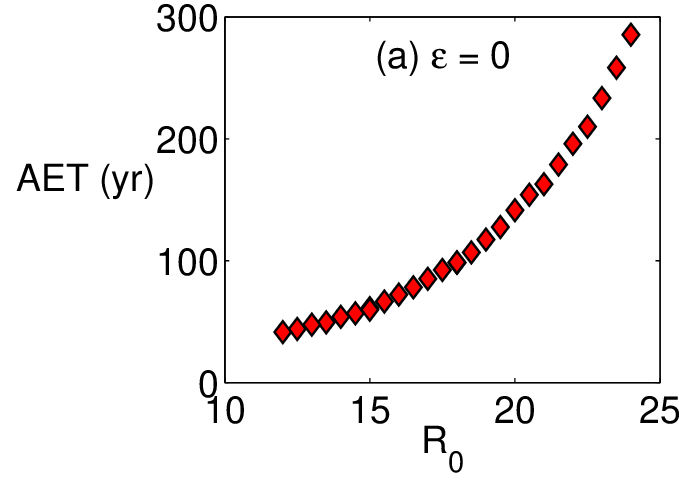}
\includegraphics[trim=0cm 0cm 0cm 0cm, clip=true, width=0.629\textwidth]{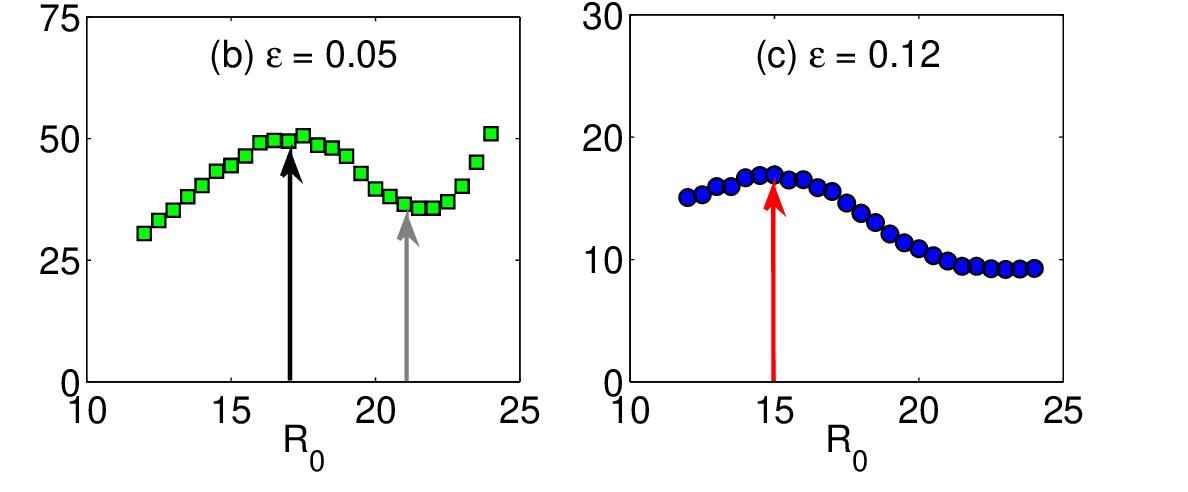}
{\caption{AET (measured in years) from simulations for the 1-city SIR model with $N=4\times10^5$ (results for other population sizes are given in the SM, Figure S1). The arrows highlight the parameter regions where the attractor of the forced system changes from annual to biennial (black and red arrows), and from biennial to annual (grey arrow).
}}
\label{fig1}
\end{figure*}
\begin{figure}
\centering
\includegraphics[trim=0cm 0cm 0cm 0cm, clip=true, width=0.5\textwidth]{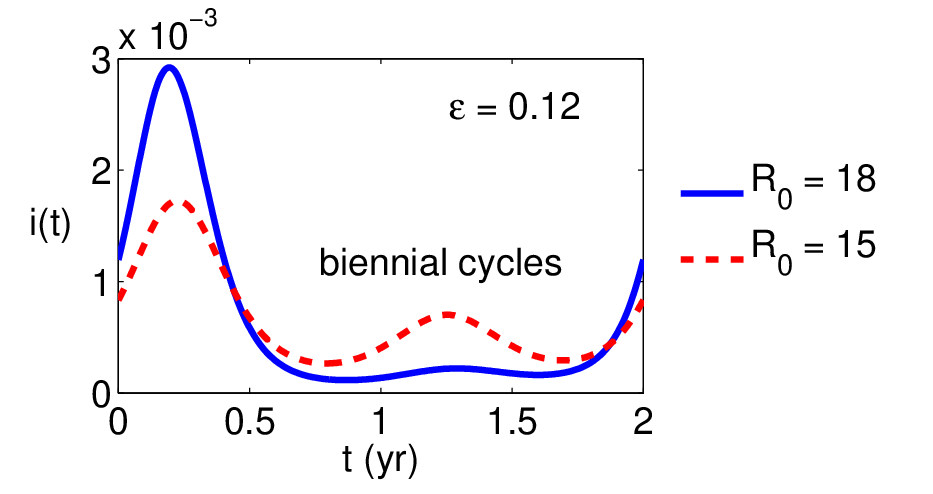}
{\caption{Fraction of infected as a function of time on the deterministic biennial attractor for the 1-city SIR model.}}
\label{fig2}
\end{figure}

To illustrate the nontrivial interplay between disease transmissibility and amplitude of seasonal forcing we start in this section by studying the stochastic SIR model in a single city. The results are shown in Figure 1. For $\epsilon=0$, the AET increases exponentially with $R_0$, but a more complex dependence on $R_0$  is found when $\epsilon>0$. More specifically, for $\epsilon=0.05$ the AET gets larger as we increase $R_0$ from 12 to 17 where it attains a maximum. A further increase of $R_0$ reduces the AET, which reaches a minimum at $R_0\approx21$ and starts to increase again from then on. For stronger seasonality, $\epsilon=0.12$, the curve describing the dependence of the AET on $R_0$ has only one maximum at about $R_0=15$.  A similar effect  was found in \cite{conlanBirth:2007} for the dependence of persistence on birth rate. The dependence of the AET on $R_0$ is even more pronounced for larger population sizes, while it becomes less significant for smaller population sizes where frequent extinctions dominate all regimes (see Section 4 of Supplementary Material). 

As in \cite{Mantilla:2010}, a simple qualitative explanation of the behavior seen in Figure 1 can be given in terms of attractors of the deterministic model and their stability properties. For the unforced case, the attractor is a stable fixed point with nonzero densities of susceptible and infective individuals. As $R_0$ increases, the infective density associated with this attractor increases, as well as its stability. As a consequence, in the stochastic model the relative amplitude of the oscillations of the number of infected around their average value gets smaller, causing the AET to increase with $R_0$. In the presence of seasonality, $\epsilon>0$, the attractors of the system are limit cycles with periods multiples of 1 year. Depending on the values of $R_0$, only annual or biennial cycles are observed in the parameter ranges we explored. Specifically, for $\epsilon=0.12$ and $R_0\in[12,14.5]$ or $R_0\in[15,24]$, the cycles are annual or biennial, respectively. The period doubling bifurcation occurs in the interval $14.5\leq R_0 \leq15$ that separates the regions of increasing and decreasing extinction times. As shown in Figure 2, the deterministic biennial cycle changes with increasing $R_0$ so that the infective density stays low for a longer fraction of the period. Due to this, in the biennial regime the AET decreases with $R_0$. In the annual regime, the stability of the cycle and the infective density averaged over the cycle increase with $R_0$, explaining the increasing AET for $R_0\in[12,14.5]$ as in the unforced case. A similar analysis holds for $\epsilon=0.05$.

\begin{figure*}
\centering
\includegraphics[trim=0cm 0cm 0cm 0cm, clip=true, width=0.707\textwidth]{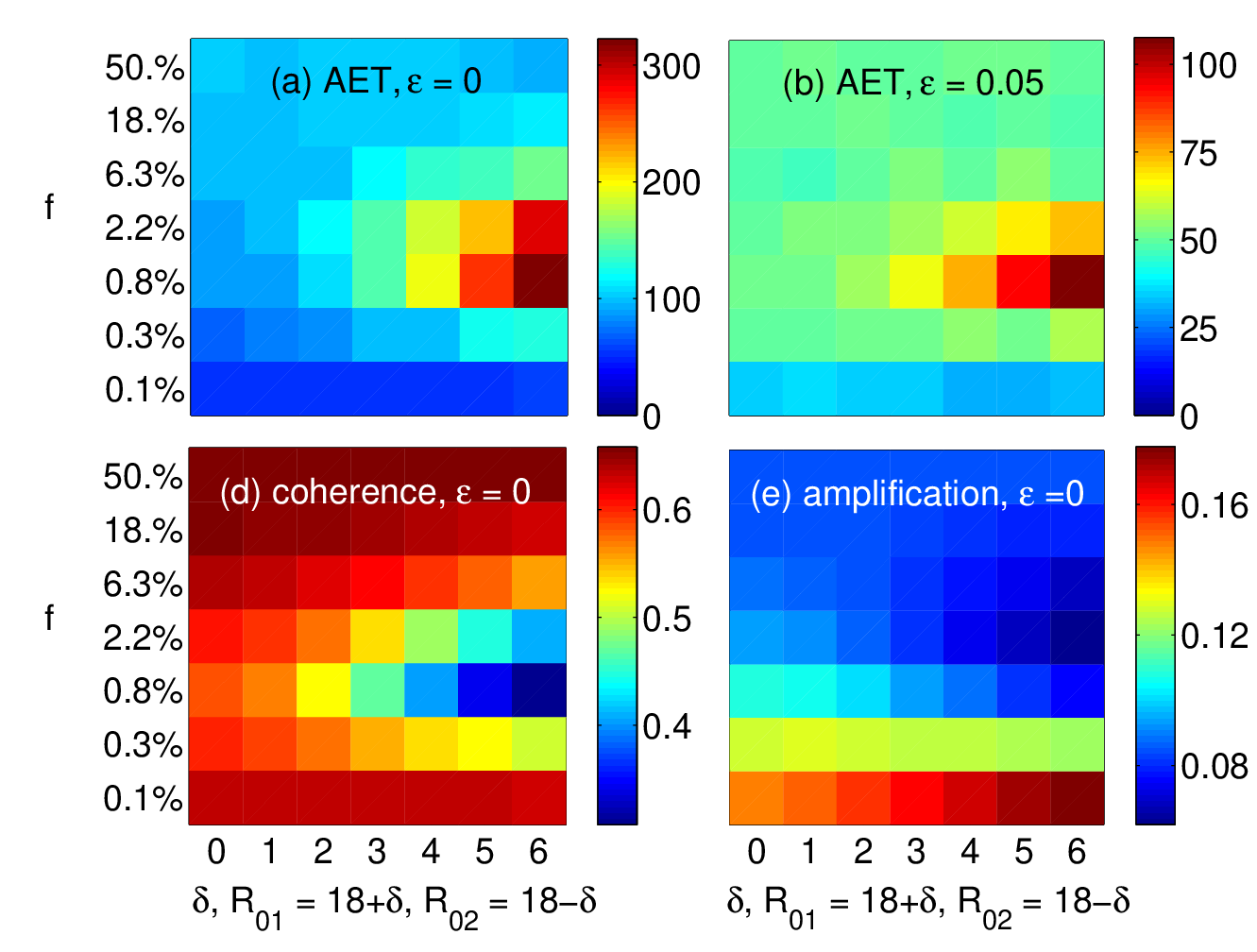}
\includegraphics[trim=0cm 0cm 0cm 0cm, clip=true, width=0.283\textwidth]{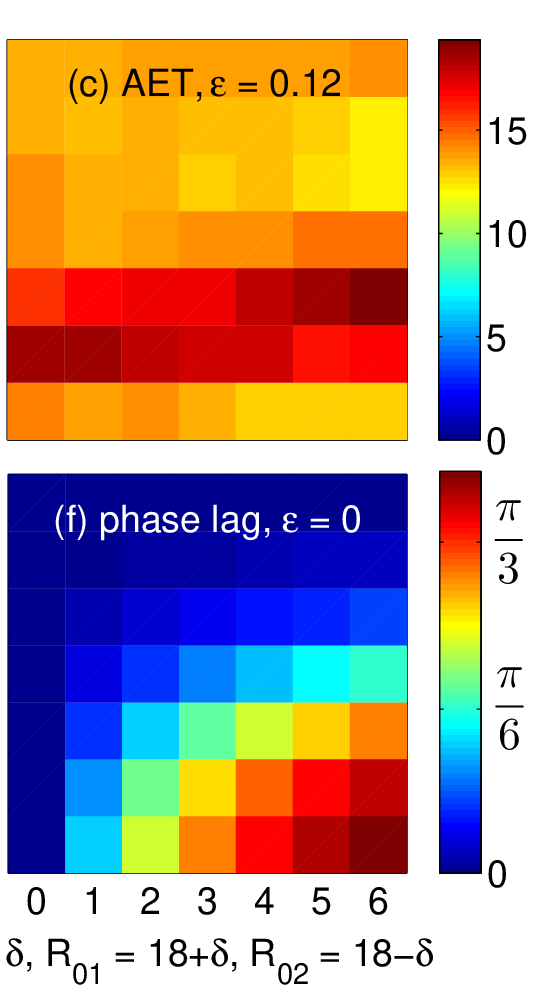}
{\caption{Results for the 2-city SIR model with $N_1=N_2=2\times10^5$. In all panels, the x-axis represents the heterogeneity in terms of $\delta = (R_{01} - R_{02})/2$ and the why axis the coupling in terms of $f$. (a)-(c) AET (measured in years) as a function of the heterogeneity parameter $\delta $ and the coupling coefficient $f$ obtained  from simulations for different levels of seasonality $\epsilon$ (notice the logarithmic scale in the $f$ axis). (d)-(f) Analytical results for the amplification and coherence in one of the cities (similar results for the other city are shown in the SM, Figure S2) and phase lag between cities for $\epsilon=0$.}}
\label{fig3}
\end{figure*}
\begin{figure*}
\centering
\includegraphics[trim=0cm 0cm 0cm 0cm, clip=true, width=0.513\textwidth]{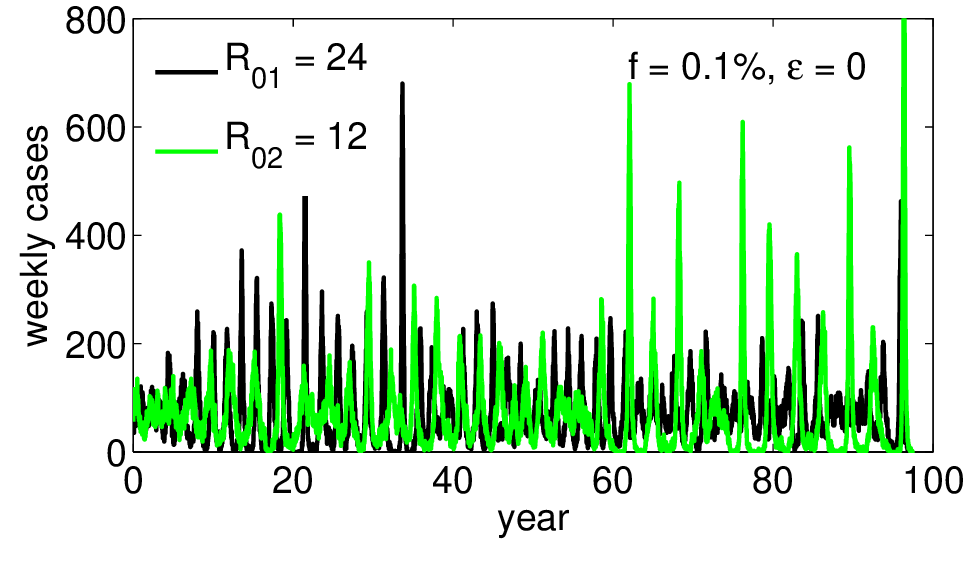}
\includegraphics[trim=0cm 0cm 0cm 0cm, clip=true, width=0.477\textwidth]{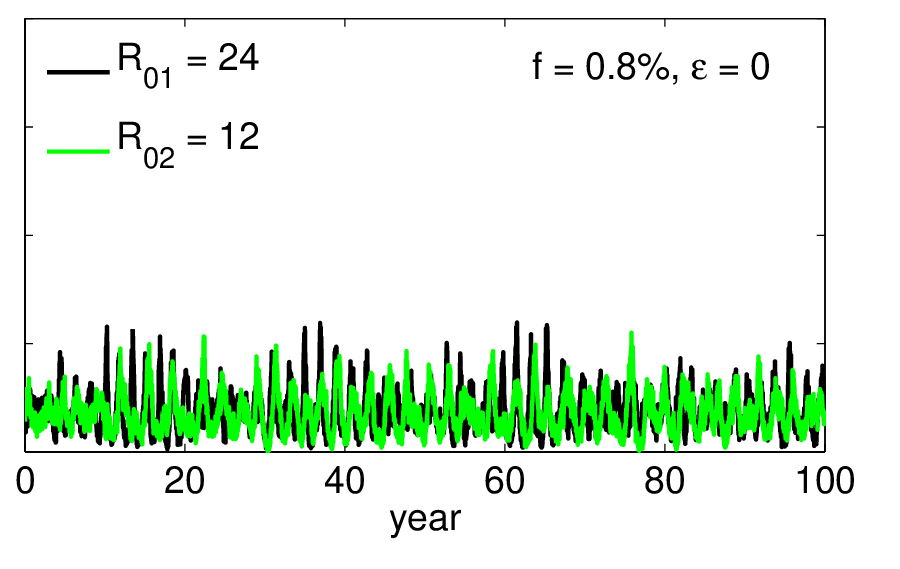}
{\caption{Typical time series of new disease cases per week for the 2-city SIR model with $N_1=N_2=2\times10^5$ and $\epsilon=0$.}}
\label{fig4}
\end{figure*}

\begin{figure*}[t]
\centering
\includegraphics[trim=0cm 0cm 0cm 0cm, clip=true, width=0.403\textwidth]{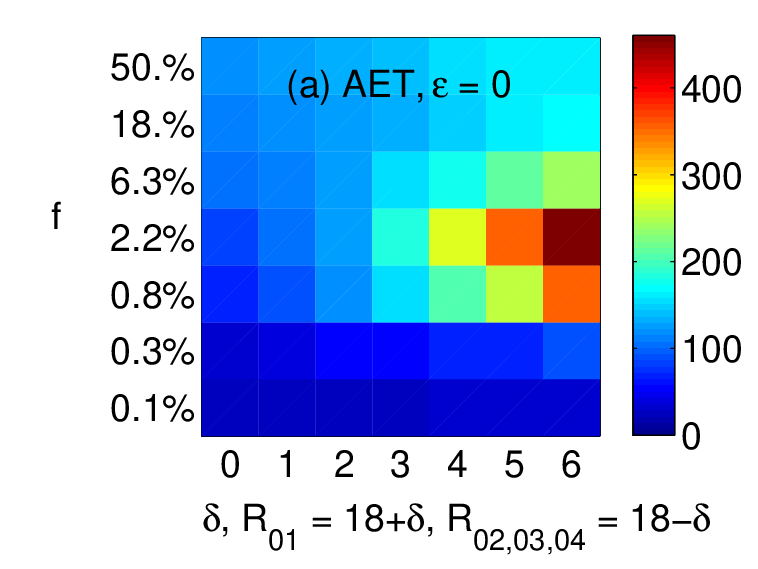}
\includegraphics[trim=0cm 0cm 0cm 0cm, clip=true, width=0.587\textwidth]{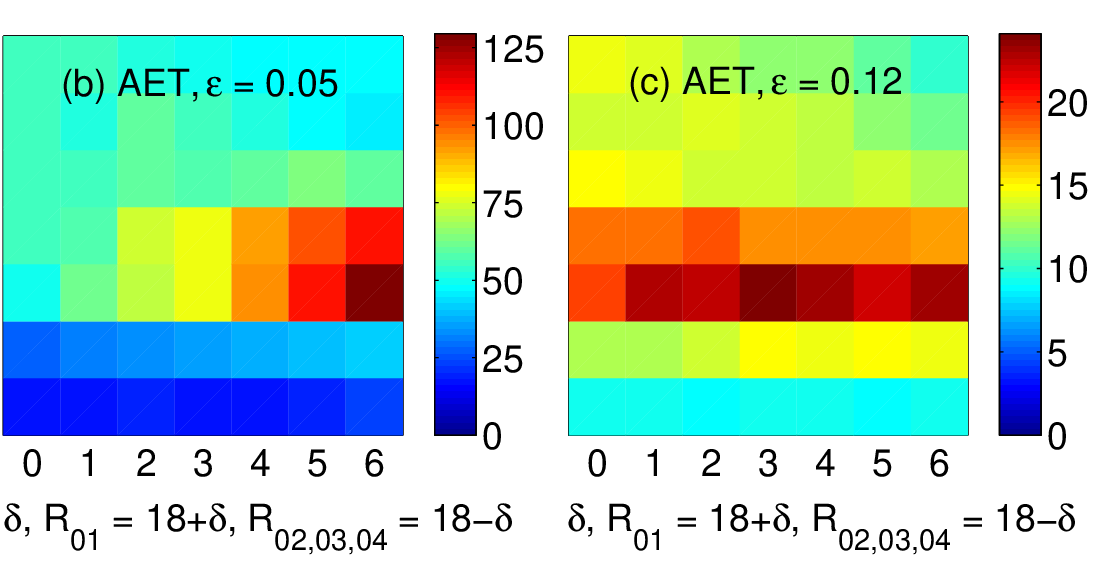}
{\caption{AET (measured in years) from simulations for the 4-city SIR model in which a central city with $N_1=2.1\times10^5$ is surrounded by 3 non-interacting satellite cities with $N_2=N_3=N_4=N_1/3$. Fraction of commuters from each of the suburban cities to the central city, $f$, is 10 times larger than that from the central city to that suburban city. In all panels, the x-axis represents the heterogeneity in terms of $\delta = (R_{01} - R_{02})/2$. Analytical results for the amplification, coherence and phase lag of the fluctuations are given in the SM, Figure S3.}}
\label{fig5}
\end{figure*}

We now explore the situation where different population patches represent very different urban environments, as may be the case of an active city centre and its suburban satellite towns. We will therefore consider our model for two cities with different values of $R_0$, taking for simplicity equal population sizes and the symmetric coupling described in Section \ref{sec:model}. The two values $R_{01}$ and $R_{02}$ that characterize SIR dynamics in each patch are given in terms of a single parameter $\delta$ as $18 \pm \delta$, thus ensuring that the average disease infectiousness is kept constant as the spatial heterogeneity increases with $\delta$. The qualitative results are independent of this assumption (results not shown), which however reduces the number of free parameters and is consistent with the fact that estimates of $R_0$ come from coarse grained data. We will study the three levels of seasonality $\epsilon=0$, $0.05$, $0.12$ considered in the study of one city, and the free parameters for each value of $\epsilon$ are $\delta$ and  the coupling strength $f$, the latter ranging from $f=0$ to the maximum meaningful value of $f=1/2$.

The results found for the AET (measured in years) are shown in Figure 3, top row.  
 For the highest level of forcing, $\epsilon=0.12$, AETs are very short and the dependence on $f$ and $\delta$ is weak, with only a slight increase in persistence for small coupling strengths, independently of heterogeneity. This is in line with results reported in the literature for uniform systems \cite{Hanski:1999}, and can be understood as temporal heterogeneity superseding spatial heterogeneity for this level of seasonality, bringing the system close to extinction for a large part of the year.

The picture that emerges for $\epsilon=0$ and $\epsilon=0.05$ is significantly different. Extinction events are rare, and when spatial heterogeneity, measured by $\delta$, is small, the AET increases with the coupling strength $f$. This is to be expected, because disease that goes extinct in one of the identical patches is more likely to be reignited by surviving infectives in the other patch as $f$ increases. More surprisingly, a second and dominant effect of persistence enhancement shows up for intermediate coupling strengths when the degree of spatial heterogeneity is high.

In order to understand the causes of this remarkable increase in persistence, we have analyzed the amplification, the coherence and the phase difference for $\epsilon=0$ and all the parameter values used to compute the AETs. We have used the analytic approximations derived in the SM, which for the population sizes under consideration provide very good quantitative agreement with simulations.
The results are shown in Figure 3, bottom row. In the left and middle panels, it can be seen that the enhanced persistence effect is associated with a reduction of the overall power of the infective number fluctuations and, more evidently, with a pronounced reduction of their coherence. In the right panel, we see that the phase lag between patches introduced by spatial heterogeneity, which would seem a good candidate to explain the observed effect, has no significant bearing on persistence.

In Figure 4 we plot two typical time series taken for $\delta=6$ and values of $f$ that correspond to extreme values of amplification and coherence in Figure 3. These plots illustrate how the amplification and coherence measures translate into the amplitude and structure of the fluctuations of the infective time series, and therefore into persistence as well.  It can be seen that in the right panel the overall amplification is more evenly distributed over frequencies, so that large amplitude fluctuations are absent and local persistence is increased. Using the theory developed in \cite{rozhnova:12b}, we show in the SM how this effect is associated with the dependence on $f$ and $\delta$ of the real part of one of the eigenvalue pairs of the linear approximation of the deterministic system at the endemic equilibrium. 

Finally, in Figure 5 we show a plot of AETs similar to that of Figure 3, but in this case for a configuration of four population patches, one central city connected to three non-interacting satellite suburban areas with equal population sizes. The results shown correspond to the central city and to one of the suburban areas for the case when the fraction of commuters from each of the suburban cities to the central city, $f$, is 10 times larger than that from the central city to that suburban city.  Both the phenomenological description and the theoretical interpretation given above for two cities hold in this case as well. The overall larger AETs in this case are due to the assymetric coupling, which increases the effective population size of the central city.

\section{Discussion}

Using a stochastic SIR metapopulation model with one population patch connected to one or several non-interacting patches, we have explored the combined effects on disease persistence of seasonality and spatial heterogeneity. The parameters in this analysis are the level of seasonality, the strength of the coupling, and the degree of heterogeneity measured by the difference between patches in the rates of potentially infectious contacts. We have considered only the simplest metapopulation structures and couplings. Results in more general settings (not shown) indicate that the effects described in this paper are robust with respect to different choices of these interaction parameters, and also to changes in the average $R_0$. In order to understand the mechanisms at play, we have compared for a large set of parameter choices the AETs found in the simulations with three main properties of the infective fluctuations: the amplitude, that measures the overall power, the coherence, that measures their regularity, and the phase lag between cities. These were computed for the unforced system  using an analytic approximation developed in \cite{rozhnova:12b}.

In contrast with spatially structured systems formed by similar patches \cite{Lloyd:1996}, 
the inherent heterogeneity of the model has been shown to induce well defined phase lags in the epidemic bursts that take place in different patches \cite{rozhnova:12b}, suggesting a simple mechanism through which heterogeneity might contribute to an increase in disease persistence. Indeed, in-phase abundance oscillations in different patches
are often associated with global extinction \cite{Kleczkowski:1995,Blasius:1999,grassly:06}.

However, we have found no clear evidence of such relation. This negative result can be understood because epidemic bursts come in short spikes, after which the system remains for a relatively long time close to extinction. The overall duration of the regime characterized by very low population numbers in two interacting patches is only slightly reduced by the phase lag.  We speculate that in systems where the population fluctuations are smoother a relation between persistence and the phase lags due to spatial heterogeneity would be apparent.

We find instead a remarkable effect of enhanced persistence associated with strong spatial heterogeneity, intermediate coupling strength and moderate seasonal forcing. We should point out that the figures illustrating this effect are shown in logarithmic scale for the coupling strength $f$, and that therefore the range of values of $f$ that produce enhanced persistence is quite small. For these values however the extinction times raise very significantly, both for the unforced and the seasonally forced system, provided the forcing amplitude is not too large.

Enhanced persistence for intermediate coupling strengths has been reported for uniform metapopulation models 
\cite{keeling:2000,Jesse:2011}. In these systems, the parameter values are the same in different patches, and spatial heterogeneity, instead of being built in the model, shows up 
only when the coupling strength is small enough for the dynamics in different patches
to be practically uncorrelated. The increase of persistence at intermediate coupling expresses the trade-off between heterogeneity due to patch structure, which is lost at high coupling, and rescue events, which become negligible at low coupling. 

One may try to carry this simple explanation over to the two patch model considered here. When the parameter values are similar in the two patches, the AET increases with the coupling strength and there is no phase lag between patches, suggesting that rescue effects are dominant in this regime. For highly heterogeneous patches, a phase lag shows up that decreases with coupling strength, while the number of rescue events increases, and the balance of these  effects would explain the increase of persistence at intermediate coupling.

In this interpretation, the increase in persistence is a consequence of the patch structure, and not of a change in the properties of the fluctuations in each patch. 
However, comparison of the AETs with the properties of the fluctuation spectrum shows that, for the unforced system, enhanced global persistence is associated with a decrease in the coherence of the fluctuations in both patches and therefore with enhanced local persistence. As detailed in the SM, the analytic approximation can be used to show that this effect is related with an increase in the stability of the attractor that describes the system for very large population sizes. This analysis was carried out for the unforced case only, but we found the effect of enhanced persistence to be relatively insensitive to seasonal forcing, provided the forcing amplitude is not so strong that it drives the system close to extinction for a large period of the year \cite{grassly:06}. We conclude that the effect of enhanced persistence documented here has to be traced back to the dependence of the stability of the attractor of the system on the coupling strength. Rather than being the result of rescue effects between population patches, it reflects an increase in local persistence induced by the coupling.

{\vskip 5pt}
{\small Financial support from the Portuguese Foundation for Science and Technology (FCT) under Contracts 
POCTI/ISFL/2/261 and (G.R.) SFRH/BPD/69137/2010  is gratefully acknowledged.} 

\bibliographystyle{unsrt}

\end{document}


\bigskip

\centerline{\Large \textbf{Supplementary Material}}

\vspace{1cm}

{\large{\bf Impact of commuting on disease persistence in heterogeneous}}

\centerline{\large{\bf  metapopulations}}

\vspace{0.7cm}

\centerline{\large{\textit{G.~Rozhnova, A.~Nunes and A.~J.~McKane}}}

\vspace{0.4cm}

In this supplementary material we will specify the model we are studying and
outline the methods we use to study it. In the final section we will also give
some additional results. Our aim in the first three sections is to give a 
non-technical introduction to the methods we use; technical descriptions 
already exist in several of our published papers [1-4]. In addition, [5] gives 
a non-technical review of stochastic modelling and [6] a more mathematical 
treatment of the same topic. We will explain the formalism in detail for the 
usual (one-city) model in such a way that the generalisation to two, or $n$, 
cities is straightforward.

\section{The Stochastic SIR model}
\subsection{Formulation of the individual-based model and the master equation}
We begin with the simple SIR model --- or in the context of this paper, the 
one-city SIR model. We define it in terms of ``reactions'' between the $N$ 
constituents of the system, which in this case are $N$ individuals which 
are either of type $S$ (susceptible), type $I$ (infected) or type $R$ 
(recovered). These individuals are born and die at the same rate, and in 
addition these events are linked, so that the total population $N$ remains 
constant. Thus instead of having distinct birth and death events, only 
combined birth/death events occur. Since all newly-born individuals are 
susceptible, these events correspond to an infected individual being replaced 
by a susceptible individual or a recovered individual being replaced by a 
susceptible individual. A susceptible individual dying results in no change in
the state of the system, since they are simply replaced by another susceptible 
individual.

The reactions that define the one-city SIR model are then:
\begin{equation}
\begin{array}{l r c l}
\textrm{1. Infection}   & \quad S+I & \stackrel{\beta}{\longrightarrow}    
& I +I  \\
\textrm{2. Recovery} & \quad I  & \stackrel{\gamma}{\longrightarrow}  & R  \\
\textrm{3. Birth/death} & \quad I & \stackrel{\mu}{\longrightarrow} 
& S  \\
\textrm{4. Birth/death} & \quad R & \stackrel{\mu}{\longrightarrow} 
& S. 
\end{array}
\label{rates}
\end{equation}
Here $\beta, \gamma$ and $\mu$ are respectively the rate of infection, the 
rate of recovery and the rate of birth/death. The reactions will be labelled by
$\alpha$ ($=1,2,3,4$).

Frequently, birth and death processes are assumed to happen at the same rate, 
but remain as distinct events. This still results in fluctuations in the total 
population size, even if the average population size remains constant. By 
linking these events at the stochastic level, the population size remains 
constant at any system size, so that we can still eliminate the variable 
relating to recovered individuals: $R=N-I-S$. This means that there are only
two variables which define the state of the system: the number of susceptible 
individuals at a given time which we will denote by $S$ and the number of 
infected individuals at a given time which we will denote by $I$. The use of
the same symbol for an individual of a particular type and the number of 
individuals of that type should cause no confusion. The general state will
then be denoted by $\mathbf{n} = (S,I)$.

The four processes (\ref{rates}) in the SIR model which cause transitions to 
a new state have the following transition rates:
\begin{equation}
\begin{array}{l r l}
\textrm{1. Infection} & T_1(S-1,I+1|S,I)\ = & \beta S I/N \\
\textrm{2. Recovery}  & T_2(S,I-1|S,I)\ = & \gamma I \\
\textrm{3. Birth/death} & T_3(S+1,I-1|S,I)\ = & \mu I \\
\textrm{4. Birth/death} & T_4(S+1,I|S,I)\ = & \mu\left( N - S - I \right).
\end{array}
\label{t_rates}
\end{equation}
We use the convention whereby the initial state is on the right and the final
state is on the left, so that $T(\mathbf{n}|\mathbf{n}')$ represents the 
transition rate from $\mathbf{n}'$ to $\mathbf{n}$.

The transition rates (\ref{t_rates}) define the individual-based SIR model,
and when substituted in the master equation (Eq.~(1) of the main text), give 
an equation for the probability, $P_{\mathbf{n}}(t)$, of finding the system in 
the state $\mathbf{n}$ at time $t$. Since the master equation can only be 
solved for very simple linear systems, we require an approximation scheme to 
make progress. Before we discuss this, we first derive the macroscopic 
description of the SIR model (in the form of differential equations) from
the master equation.

To do this it is useful to write the master equation in a slightly different 
form, by introducing the \textit{stoichiometric coefficients} 
$\boldsymbol{\nu}_{\alpha}$, for reaction $\alpha$. They indicate by how much 
$S$ or $I$ increase or decrease in a given reaction, that is, 
$\mathbf{n} = \mathbf{n}' + \boldsymbol{\nu}_\alpha$. So for instance in 
reaction 1, the components of $\boldsymbol{\nu}_{1}$ are 
$\nu^{(1)}_1 = -1, \nu^{(2)}_1 = +1$, and for reaction 2 we have 
$\nu^{(1)}_2 = 0, \nu^{(2)}_2 = -1$, and so on. Here the superscript $(1)$ refers
to $S$ and $(2)$ to $I$. Then the master equation can be written as 
\begin{equation}
\frac{\mathrm{d}P_{\boldsymbol{n}}(t)}{\mathrm{d}t} = \sum^{M}_{\alpha=1}
\left[ T_{\alpha}(\boldsymbol{n}|\boldsymbol{n}-\boldsymbol{\nu}_{\alpha})
P_{\boldsymbol{n}-\boldsymbol{\nu}_{\alpha}}(t) - 
T_{\alpha}(\boldsymbol{n}+\boldsymbol{\nu}_{\alpha}|\boldsymbol{n})
P_{\boldsymbol{n}}(t) \right],
\label{master_alt}
\end{equation}
where in this case $M=4$.

\subsection{The macroscopic equation}
The macroscopic equation is found by multiplying Eq.~(\ref{master_alt}) by 
$\boldsymbol{n}$, and summing over all possible values of $\boldsymbol{n}$. 
After making the change of variable $\boldsymbol{n} \rightarrow
\boldsymbol{n}+\boldsymbol{\nu}_\alpha$ in the first summation, one finds that
\begin{equation}
\frac{\mathrm{d}\langle \boldsymbol{n}(t) \rangle}{\mathrm{d}t} = 
\sum^{M}_{\alpha=1} \boldsymbol{\nu}_{\alpha} 
\big\langle T_{\alpha}(\boldsymbol{n}+\boldsymbol{\nu}_{\alpha}|\boldsymbol{n}) 
\big\rangle,
\label{av_n_eqn}
\end{equation}
where the angle brackets define the expectation value:
\begin{equation}
\langle\cdots\rangle = \sum_{\bm{n}}(\cdots)P_{\bm{n}}(t)\,.
\end{equation} 

The first approximation we will make (which is exact in the limit 
$N \to \infty$) is to label the states of the system by the fraction of the 
population which are susceptible and infected, rather than the numbers of 
susceptible and infected individuals. In other words, we will use $s=S/N$ and 
$i=I/N$ (in the limit $N \to \infty$) rather than $S$ and $I$, and treat these 
as continuous variables. This is the sometimes called the diffusion 
approximation. So the state of the system will be determined by the new 
variables 
\[
\boldsymbol{\phi}= (s,i) = 
\lim_{N \to \infty} \frac{\langle \mathbf{n} \rangle}{N}.
\]
Dividing Eq.~(\ref{av_n_eqn}) by $N$, we have that
\begin{equation}
\frac{\mathrm{d}\boldsymbol{\phi}}{\mathrm{d}t}=
\sum_{\alpha=1}^M \boldsymbol{\nu}_{\alpha}g_{\alpha}(\boldsymbol{\phi}),
\label{ODEs_gen}
\end{equation}
where the function $g_{\alpha}(\boldsymbol{\phi})$ is defined by
$\lim_{N \to \infty} N^{-1}\,\langle T_{\alpha}(\bm{n}+
\boldsymbol{\nu}_{\alpha}|\bm{n}) \rangle$.

Equation (\ref{ODEs_gen}) is the usual deterministic (macroscopic) equation 
for the SIR model, and specifies how the mean fraction of susceptible and 
infected individuals changes with time. To write it in a more familiar form, 
we first note that in the limit $N \to \infty$ the system is deterministic, 
and so the variance (and higher cumulants) of $\bm{n}$ are zero. This implies \
that 
\[
\lim_{N \to \infty} N^{-1}\,\langle T_{\alpha}(\bm{n}+
\boldsymbol{\nu}_{\alpha}|\bm{n}) \rangle = \lim_{N \to \infty} 
N^{-1}\,T_{\alpha}\big(\langle\bm{n}\rangle+
\boldsymbol{\nu}_{\alpha}|\langle\bm{n}\rangle\big).
\]
The mean value $\langle\bm{n}\rangle$ can now simply be replaced by
$N\boldsymbol{\phi}$, leading to the identification:
\begin{equation}
g_{\alpha}(\boldsymbol{\phi}) = \lim_{N \to \infty} 
N^{-1}\,T_{\alpha}(N\boldsymbol{\phi}+
\boldsymbol{\nu}_{\alpha}|N\boldsymbol{\phi})\bigg.\,.
\label{f_defn}
\end{equation}
For example, from Eq.~(\ref{t_rates}), $T_1(S-1,I+1|S,I)=\beta S I/N$, and
so $g_1 = \lim_{N \to \infty} N^{-1}\,T_{1}=\beta s i$. This leads to the 
following characterisation of the reactions --- which is the only information 
we require for specifying both the deterministic and stochastic dynamics of 
the system:
\begin{equation}
\begin{array}{l l l}
\textrm{1. Infection} & \nu^{(1)}_1 = -1, \ \nu^{(2)}_1 = +1 & \ \ \ 
g_1 = \beta s i \\
\textrm{2. Recovery}  & \nu^{(1)}_2 = 0, \ \ \ \nu^{(2)}_2 = -1 & \ \ \
g_2 = \gamma i \\
\textrm{3. Birth/death} & \nu^{(1)}_3 = +1, \ \nu^{(2)}_3 = -1 & \ \ \ 
g_3 = \mu i \\
\textrm{4. Birth/death} & \nu^{(1)}_4 = +1, \ \nu^{(2)}_4 = 0 & \ \ \ 
g_4 = \mu\left( 1 - s - i \right).
\end{array}
\label{info_req}
\end{equation}

We define 
\begin{equation}
\mathbf{A}(\boldsymbol{\phi})=
\sum_{\alpha=1}^M \boldsymbol{\nu}_{\alpha}g_{\alpha}(\boldsymbol{\phi}),
\label{A_def}
\end{equation}
so that the macroscopic equation (\ref{ODEs_gen}) is 
\begin{equation}
\frac{\mathrm{d}\boldsymbol{\phi}}{\mathrm{d}t}=\mathbf{A}(\boldsymbol{\phi}).
\label{macro_eqn}
\end{equation}
Writing this in terms of the components of 
$\boldsymbol{\phi}=\left( s,i \right)$ we find, using Eq.~(\ref{info_req}), 
the well-known equations for the deterministic SIR model:
\begin{eqnarray}
\frac{\mathrm{d}s}{\mathrm{d}t} &=& - \beta si + \mu i +
\mu\left( 1 - s - i \right), \nonumber \\
\frac{\mathrm{d}i}{\mathrm{d}t} &=& + \beta si - \gamma i - \mu i.
\label{SIR_deter}
\end{eqnarray}
As is also well-known these equations have two fixed points: one where there
are no infected individuals in the population, that is, $(s^*,i^*)=(1,0)$, and
the one which is of most interest:
\begin{equation}
s^* = \frac{\gamma + \mu}{\beta}, \ \ \ \ i^* = \frac{\mu\left[ \beta 
- (\gamma + \mu) \right]}{\beta(\gamma + \mu)}.
\label{FP}
\end{equation}
In the above the asterisk denotes a fixed point of the differential equations. 

\subsection{The stochastic equation}
A question of immediate interest which can be answered from the 
deterministic equation (\ref{SIR_deter}) is the long term behaviour of the 
model, which involves finding the stability of the fixed points. We carry 
out this analysis in the usual way, by linearising the equations 
(\ref{SIR_deter}) about the fixed point. That is, we substitute 
\begin{equation}
s = s^* + \frac{x}{\sqrt{N}}, \ \ \  i = i^* + \frac{y}{\sqrt{N}},
\label{linearisation}
\end{equation}
into the equations (\ref{SIR_deter}), keeping only linear terms in either $x$
or $y$. This leads to the equations
\begin{eqnarray}
\frac{\mathrm{d}x}{\mathrm{d}t} &=& {\mathcal J}_{11}x + 
{\mathcal J}_{12}y, \nonumber \\
\frac{\mathrm{d}y}{\mathrm{d}t} &=& {\mathcal J}_{12}x + 
{\mathcal J}_{22}y,
\label{LSA}
\end{eqnarray}
where the Jacobian ${\mathcal J}$ is evaluated at the non-trivial fixed point
\begin{equation}
{\mathcal J} = \left( 
\begin{array}{cc}
-\beta i^* - \mu & -\beta s^* \\ \\
\beta i^* & \beta s^* - (\gamma + \mu)
\end{array} 
\right) = \left( 
\begin{array}{cc}
-\frac{\beta\mu}{\gamma + \mu} & - (\gamma + \mu) \\ \\
\frac{\left[ \beta - (\gamma + \mu)\right]}{\beta} & 0
\end{array} 
\right).
\label{Jacobian} 
\end{equation}
The eigenvalues of this matrix have a negative real part, and are in fact a 
complex conjugate pair for all realistic values of the rates $\beta, \gamma$ 
and $\mu$. Thus this fixed point is stable and small perturbations undergo 
damped oscillations towards $(s^*,i^*)$.

The stochastic version of the model can now be easily written down. We will only
be interested in fluctuations about the stationary state of system, that is, 
about the fixed point (\ref{FP}). Furthermore we will only investigate 
\textit{linear} fluctuations, since numerous studies have shown that
this captures the actual dynamics very well indeed. This is known as the 
linear noise approximation (LNA) or the next-to-leading order in the van
Kampen system-size expansion. It consists of implementing the expansion
(\ref{linearisation}) but on the master equation, not on the deterministic 
equations. The latter calculation, described above, amounted to an analysis 
of what happened to a single small perturbation to the deterministic dynamics.
The linearisation carried out on the master equation, in contrast, results in 
extra ``noise'' terms which are added to (\ref{LSA}) and which continuously 
modify the (stochastic) perturbation. In this context the factors of $N^{-1/2}$
in the linearisation (\ref{linearisation}) are important; they encode the 
fact that fluctuations will naively be of this order, and so pick out the 
linear correction in the expansion procedure. The $N^{-1/2}$ factors were
included in the linear stability analysis of the deterministic equation to
highlight the connection with the LNA, but clearly have no effect on 
Eq.~(\ref{LSA}), since any factor multiplying the perturbation will cancel 
out in this linear equation.

We will not carry out the calculations leading to the equations describing the
stochastic fluctuations about the fixed point, since they have already 
appeared several times in the literature in various forms. The equations are
\begin{eqnarray}
\frac{\mathrm{d}x}{\mathrm{d}t} &=& {\mathcal J}_{11}x + 
{\mathcal J}_{12}y + \eta_1(t), \nonumber \\
\frac{\mathrm{d}y}{\mathrm{d}t} &=& {\mathcal J}_{12}x + 
{\mathcal J}_{22}y + \eta_2(t),
\label{LNA}
\end{eqnarray}
where $\eta_1(t)$ and $\eta_2(t)$ are noises with a Gaussian distribution and
zero mean, and with a correlation function given by
\begin{equation}
\langle \eta_J(t) \eta_K(t') \rangle = B_{JK} \delta(t-t').
\label{white_noise}
\end{equation}
Here $B_{JK}$ is given by
\begin{equation}
B_{JK}(\boldsymbol{\phi})=
\sum_{\alpha=1}^M 
\nu^{(J)}_{\alpha}\,\nu^{(K)}_{\alpha}\,g_{\alpha}(\boldsymbol{\phi}),
\ \ \ \ J,K=1,2,
\label{B_def}
\end{equation}
evaluated at the fixed point (\ref{FP}). Using Eqs.~(\ref{info_req}) and 
(\ref{FP}) one finds that
\begin{equation}
B = \left( 
\begin{array}{cc}
\beta s^* i^* +\mu(1 - s^*) & -\beta s^* i^* -\mu i^* \\ \\
-\beta s^* i^* -\mu i^* & \beta s^* i^* + (\gamma + \mu)i^*
\end{array} 
\right) = \left( 
\begin{array}{cc}
\frac{2\mu\,\left[ \beta - (\gamma + \mu)\right]}{\beta} & 
-\frac{\mu (\gamma + 2\mu)\,\left[ \beta - (\gamma + \mu)\right]}
{\beta(\gamma +\mu)} \\ \\
-\frac{\mu (\gamma + 2\mu)\,\left[ \beta - (\gamma + \mu)\right]}
{\beta(\gamma +\mu)} & \frac{2\mu\,\left[ \beta - (\gamma + \mu)\right]}{\beta}
\end{array} 
\right).
\label{B_JK} 
\end{equation}
Equations (\ref{LNA}), (\ref{white_noise}) and (\ref{B_JK}) completely specify 
the stochastic fluctuations about the deterministic SIR model.

To analyse the stochastic differential equations (\ref{LNA}) further it is
convenient to take their Fourier transforms and introduce
\begin{equation}
\tilde{x}(\omega) = \int^{\infty}_{-\infty} e^{i\omega t}\,x(t)\,dt, \ \ \
\tilde{y}(\omega) = \int^{\infty}_{-\infty} e^{i\omega t}\,y(t)\,dt.
\label{FT}
\end{equation}
The lower limit of the integration comes from the fact that to ensure that the
system is in a stationary state, the initial conditions have to be set in the
infinitely distant past. Since the Fourier transform of 
$\mathrm{d}x/\mathrm{d}t$ is $-i\omega \tilde{x}$, one finds that
\begin{eqnarray}
\left[ -i\omega - {\mathcal J}_{11} \right] \tilde{x}(\omega) - 
{\mathcal J}_{12} \tilde{y}(\omega) &=& \tilde{\eta}_1(\omega), \nonumber \\
-{\mathcal J}_{21} \tilde{x}(\omega) + \left[ -i\omega - 
{\mathcal J}_{22} \right] \tilde{y}(\omega) &=& \tilde{\eta}_2(\omega).
\label{LNA_freq}
\end{eqnarray}
Defining the matrix $-i\omega \delta_{JK} - {\mathcal J}_{JK}$ to be 
$\Phi_{JK}(\omega)$, Eq.~(\ref{LNA_freq}) may be written in the more concise 
form 
$\sum^{2}_{K=1} \Phi_{JK}(\omega)\tilde{z}_{K}(\omega)=\tilde{\eta}_{J}(\omega)$,
where $z_1 \equiv x$ and $z_2 \equiv y$. This may be solved to yield
\begin{equation}
\tilde{z}_{J}(\omega) = \sum^{2}_{K=1} \Phi^{-1}_{JK}(\omega) 
\tilde{\eta}_{K}(\omega),
\label{solution}
\end{equation}
where $\Phi^{-1}$ is the inverse of the matrix $\Phi$.

\medskip
 
The nature of the stochastic fluctuations can usefully be investigated by 
calculating the power spectrum defined by 
\begin{equation}
P_{J}(\omega) \equiv \left\langle | \tilde{z}_J(\omega) |^{2} \right\rangle
= \sum^{2}_{K=1} \sum^{2}_{L=1} \Phi^{-1}_{JK}(\omega)B_{KL}
\left( \Phi^{\dag} \right)^{-1}_{LJ}(\omega),
\label{PS_defn}
\end{equation}
where $\Phi^{\dag}$ is the Hermitian conjugate of the matrix $\Phi$. Examples
of power spectra from a number of models are given in [1-4]. To characterise
different spectra we use two measures. The amplification is defined as the overall power of the fluctuation spectrum. The coherence is the fraction of the total power in a 10\% frequency range around the dominant frequency of
the spectrum.

\smallskip

The explicit form of the denominator in Eq.~(\ref{PS_defn}) is 
$|\mathrm{det} \Phi(\omega)|^2$. From this we can deduce that a pair of 
complex conjugate eigenvalues of ${\cal J}$ with small real parts will give 
rise to a peak in the power spectrum, whose location in frequency is 
determined by the imaginary parts~[3]. The file ESM.mov included as 
Supplementary Material shows how the persistence enhancement effect discussed 
in the main text is associated with the dependence of one such pair of 
eigenvalues of ${\cal J}$ on the parameters of the system.

\medskip

\section{The Two-City Stochastic SIR model}
Having discussed in detail the formalism and general procedure in the one-city
case in a way that naturally generalises, the description of the two-city case 
can be more concise. What is novel is the nature of the interactions between 
the two cities, which is what we discuss first of all.

\medskip

\subsection{Form of the interaction between the two cities}
The interaction between the two cities is only reflected in the first type of
reaction in Eq.~(\ref{rates}); the other three remain essentially unchanged 
since they involve only one individual, and so only one city.

The number of individuals in the three classes belonging to city $j$ ($j=1,2$)
will be denoted by $S_{j}, I_{j}$ and $R_j$ respectively. We again assume that 
births and deaths are coupled at the individual level, so that when an 
individual dies another (susceptible) individual is born. Therefore the number
of recovered individuals is not an independent variable: 
$R_{j}=N_{j}-S_{j}-I_{j}$, where $j=1,2$. Since we do not focus on specific 
individuals, we will not be concerned with the precise movements of 
commuters between the cities --- the frequency or duration of their commute ---
but only the fraction of the population which is away from its home city at 
any given time. For commuters from city $j$ to city $k$ we will denote this 
fraction by $f_{kj}$. It follows that the number of individuals in city $1$ is 
$M_{1}=(1-f_{21})N_{1}+f_{12}N_{2}$ and in city $2$ is
$M_{2}=(1-f_{12})N_{2}+f_{21}N_{1}$. 

We will assume that the birth/death rate and the recovery rate are the same 
in both cities, but that the infection rates are city dependent: $\beta_1$ in
city $1$ and $\beta_2$ in city $2$. When including seasonal forcing, as we are
in this study, the infection rates are time-dependent: 
$\beta_j(t)=\beta^0_j\,\left( 1 + \epsilon\,\cos\,2\pi t \right)$, where the 
time $t$ is measured in years and $\epsilon$ represents the amplitude of the 
seasonal forcing.

There are four different types of infection events. 
\begin{itemize}
\item[(i)] Infective residents of one city (say $j$) can infect susceptible 
residents of the same city. The rate for this to occur is
$\beta_j\,\left( 1 - f_{kj}\right)S_j\,\left( 1 - f_{kj}\right)I_j/M_j$.
\item[(ii)] Infective commuters can travel from city $k$ to city $j$ and infect 
susceptible residents in their home city $j$. The rate for this to occur is
$\beta_j\,\left( 1 - f_{kj}\right)S_j\,f_{jk}I_k/M_j$.
\item[(iii)] Infective residents in their home city (say $k$) can infect 
susceptible commuters from the other city ($j$). The rate for this to occur is
$\beta_k\,f_{kj}S_j\,\left( 1 - f_{jk}\right)I_k/M_k$.
\item[(iv)] Infective commuters can infect susceptible commuters away from 
their home city of $j$. The rate for this to occur is
$\beta_k\,f_{kj}S_j\,f_{kj}I_j/M_k$.
\end{itemize}

Adding these rates together we obtain the total transition rate for infection 
of $S_j$ individuals as 
\begin{equation}
\beta_{jj} \frac{S_j I_j}{N_j} + \beta_{jk} \frac{S_j I_k}{N_k},
\label{two_city_rates}
\end{equation}
where 
\begin{equation}
\beta_{jj} = \frac{\beta_j \left( 1 - f_{kj}\right)^2 N_j}{M_j}
+ \frac{\beta_{k}f_{kj}^2 N_j}{M_k}, \ \ \ 
\beta_{jk} = \frac{\beta_j \left( 1 - f_{kj} \right)f_{jk}N_k}{M_j} 
+ \frac{\beta_k f_{kj}\left( 1 - f_{jk}\right) N_k}{M_k}.
\label{beta_jk}
\end{equation}

Equation (\ref{two_city_rates}) gives us the generalisation of the infection 
rate to the two-city case. There are now eight processes rather than the four 
processes (\ref{rates}) of the one-city SIR model which cause transitions from 
one state to another. They are
\begin{equation}
\begin{array}{l r l}
\textrm{1. Infection of $S_1$} & T_1(S_{1}-1,I_{1}+1|S_1,I_1)\ = & 
\beta_{11} \frac{S_1 I_1}{N_1} + \beta_{12} \frac{S_1 I_2}{N_2} \\ \\
\textrm{2. Infection of $S_2$} & T_2(S_{2}-1,I_{2}+1|S_2,I_2)\ = & 
\beta_{21} \frac{S_2 I_1}{N_1} + \beta_{22} \frac{S_2 I_2}{N_2} \\ \\
\textrm{3. Recovery of $I_1$} & T_3(S_1,I_{1}-1|S_1,I_1)\ = & \gamma I_1 \\ \\
\textrm{4. Recovery of $I_2$} & T_4(S_2,I_{2}-1|S_2,I_2)\ = & \gamma I_2 \\ \\
\textrm{5. Birth/death in city 1} & T_5(S_{1}+1,I_{1}-1|S_1,I_1)\  = & \mu I_1 \\
\\
\textrm{6. Birth/death in city 2} & T_6(S_{2}+1,I_{2}-1|S_2,I_2)\  = & \mu I_2 \\
\\
\textrm{7. Birth/death in city 1} & T_7(S_{1}+1,I_{1}|S_1,I_1)\  = & 
\mu (N_{1}-S_{1}-I_{1}) \\ \\
\textrm{8. Birth/death in city 2} & T_8(S_{2}+1,I_{2}|S_2,I_2)\  = & 
\mu (N_{2}-S_{2}-I_{2}). 
\end{array}
\label{t_twocity_rates}
\end{equation}
Note that we have not listed all the state variables as arguments of the 
transition rates $T_{\alpha}(\cdots | \cdots)$ --- only those which are most 
relevant to the reaction under consideration.

We may now set up the master equation in the same way as for the one-city case.
Introducing $\mathbf{n} = (S_1,S_2,I_1,I_2)$, the master equation takes the
form (\ref{master_alt}) where now  $M=8$, and the reactions with the 
transition rates are given by Eq.~(\ref{t_twocity_rates}). The variables 
relevant for the deterministic and stochastic equations are $s_j=S_j/N_j$ and
$i_j=I_j/N_j$ ($j=1,2$), with $\boldsymbol{\phi}=(s_1,s_2,i_1,i_2)$. The 
analogue of Eq.~(\ref{info_req}) for two cities (although now with only the 
non-zero stoichiometric coefficients listed) is 
\begin{equation}
\begin{array}{l r l}
\textrm{1. Infection of $S_1$} & \nu^{(1)}_1 = -1, \ \nu^{(3)}_1 = +1 & 
g_1 = \beta_{11}s_1 i_1 + \beta_{12}s_1 i_2 \\ 
\textrm{2. Infection of $S_2$} & \nu^{(2)}_2 = -1, \ \nu^{(4)}_2 = +1 & 
g_2 = \beta_{21}s_2 i_1 + \beta_{22}s_2 i_2 \\ 
\textrm{3. Recovery of $I_1$} & \nu^{(3)}_3 = -1 & g_3 = \gamma i_1 \\
\textrm{4. Recovery of $I_2$} & \nu^{(4)}_4 = -1 & g_4 = \gamma i_2 \\ 
\textrm{5. Birth/death in city 1} & \nu^{(1)}_5 = +1, \ \nu^{(3)}_5 = -1 & 
g_5 = \mu i_1 \\
\textrm{6. Birth/death in city 2} & \nu^{(2)}_6 = +1, \ \nu^{(4)}_6 = -1 & 
g_6 = \mu i_2 \\
\textrm{7. Birth/death in city 1} & \nu^{(1)}_7 = +1 & g_7 = \mu(1 -s_{1}-i_{1}) 
\\
\textrm{8. Birth/death in city 2} & \nu^{(2)}_8 = +1 & g_8 = \mu(1 -s_{2}-i_{2}).
\end{array}
\label{info_req_two_cities}
\end{equation}

\medskip

\subsection{The macroscopic and stochastic equations}
The information given in Eq.~(\ref{info_req_two_cities}) is again sufficient 
to completely specify both the macroscopic and stochastic equations.

The macroscopic equation (\ref{macro_eqn}), where 
$\mathbf{A}(\boldsymbol{\phi})$ is defined by Eq.~(\ref{A_def}), when written 
down in terms of the components of 
$\boldsymbol{\phi}=\left( s_1,s_2,i_i,i_2 \right)$, is
\begin{eqnarray}
\frac{\mathrm{d}s_1}{\mathrm{d}t} &=& - \beta_{11}s_1i_1 - \beta_{12}s_1i_2
+ \mu i_1 + \mu\left( 1 - s_1 - i_1 \right), \nonumber \\
\frac{\mathrm{d}s_2}{\mathrm{d}t} &=& - \beta_{21}s_2i_1 - \beta_{22}s_2i_2
+ \mu i_2 + \mu\left( 1 - s_2 - i_2 \right), \nonumber \\
\frac{\mathrm{d}i_1}{\mathrm{d}t} &=& + \beta_{11} s_1i_1 + \beta_{12}s_1i_2
- \gamma i_1 - \mu i_1, \nonumber \\
\frac{\mathrm{d}i_2}{\mathrm{d}t} &=& + \beta_{21} s_2i_1 + \beta_{22}s_2i_2
- \gamma i_2 - \mu i_2.
\label{two_cities__deter}
\end{eqnarray}

It is known for two cities (and also for $n$ cities) that a unique
non-trivial fixed point exists which is globally stable. The Jacobian evaluated
at this fixed point is 
\begin{equation}
{\mathcal J} = \left( 
\begin{array}{cccc}
-\beta_{11}i_1^* -\beta_{12}i^*_2 - \mu & 0 & -\beta_{11}s^*_1 & -\beta_{12}s^*_1
\\ \\
0 & -\beta_{21}i^*_1 - \beta_{22}i^*_2 - \mu & - \beta_{21}s^*_2 & 
- \beta_{22}s^*_2 \\ \\
\beta_{11}i^*_1 + \beta_{12}i^*_2  & 0 & \beta_{11}s_1^* - (\gamma + \mu) & 
\beta_{12}s^*_1 \\ \\
0 & \beta_{21}i^*_1 + \beta_{22} i^*_2 & \beta_{21}s_2^* & \beta_{22}s^*_2
- (\gamma + \mu) 
\end{array} 
\right).
\label{Jacobian_two_cities} 
\end{equation}

\medskip

The stochastic fluctuations about the fixed point are obtained through a 
generalisation of Eq.~(\ref{linearisation}): 
\begin{equation}
s_j = s^*_j + \frac{x_j}{\sqrt{N_j}}, \ \ \  
i_j = i^*_j + \frac{y_j}{\sqrt{N_j}}, \ \ \ j=1,2,
\label{two_cities_linearisation}
\end{equation}
and keeping only linear terms in either $x$ or $y$. This leads to the equations
\begin{eqnarray}
\frac{\mathrm{d}x_1}{\mathrm{d}t} &=& {\mathcal J}_{11}x_1 + {\mathcal J}_{12}x_2
+ {\mathcal J}_{13}y_1 + {\mathcal J}_{14}y_2 + \eta_1(t), \nonumber \\
\frac{\mathrm{d}x_2}{\mathrm{d}t} &=& {\mathcal J}_{21}x_1 + {\mathcal J}_{32}x_2
+ {\mathcal J}_{23}y_1 + {\mathcal J}_{24}y_2 + \eta_2(t), \nonumber \\
\frac{\mathrm{d}y_1}{\mathrm{d}t} &=& {\mathcal J}_{31}x_1 + {\mathcal J}_{32}x_2
+ {\mathcal J}_{33}y_1 + {\mathcal J}_{34}y_2 + \eta_3(t), \nonumber \\
\frac{\mathrm{d}y_2}{\mathrm{d}t} &=& {\mathcal J}_{41}x_1 + {\mathcal J}_{42}x_2
+ {\mathcal J}_{43}y_1 + {\mathcal J}_{44}y_2 + \eta_4(t),
\label{LNA_two_cities}
\end{eqnarray}
which may be written in a more compact form by introducing the vector of 
fluctuations $\mathbf{z}=(x_1,x_2,y_1,y_2)$:
\begin{equation}
\frac{\mathrm{d}z_J}{\mathrm{d}t} = \sum^{4}_{K=1}{\mathcal J}_{JK}z_K
+ \eta_J(t), \ \ \ J=1,\ldots,4.
\label{SDE_two_cities}
\end{equation}
Here the $\eta_J(t)$ are noises with a Gaussian distribution and zero mean, 
and with a correlation function given by Eq.~(\ref{white_noise}). 

\medskip

The $B$ matrix (\ref{B_def}) for the two-city case, evaluated at a fixed point,
is
\begin{equation}
B =\left[\begin{array}{c|c} 
B^{(1)} & B^{(2)} \\ \hline 
B^{(3)} & B^{(4)} 
 \end{array}\right],
\label{block}
\end{equation}
where these submatrices are given by
\begin{equation}
B^{(1)} = \left( 
\begin{array}{cc}
\beta_{11}s^*_1 i_1^* +\beta_{12}s^*_1 i^*_2 + \mu(1-s^*_1) & 0 \\ \\
0 & \beta_{21}s^*_2 i^*_1 +\beta_{22}s^*_2 i^*_2 + \mu(1-s^*_2)
\end{array} 
\right),
\label{B(1)} 
\end{equation}
\begin{equation}
B^{(2)} = B^{(3)} = \left( 
\begin{array}{cc}
-\beta_{11}s^*_1 i^*_1 -\beta_{12}s^*_1 i^*_2 -\mu i^*_1 & 0 \\ \\
0 & -\beta_{21}s^*_2 i^*_1 -\beta_{22}s^*_2 i^*_2 -\mu i^*_2 
\end{array} 
\right),
\label{B(2)B(3)} 
\end{equation}
and
\begin{equation}
B^{(4)} = \left( 
\begin{array}{cc}
\beta_{11}s^*_1 i_1^* +\beta_{12}s^*_1 i^*_2 + (\gamma + \mu)i^*_1  & 0 \\ \\
0 & \beta_{21}s^*_2 i^*_1 +\beta_{22}s^*_2 i^*_2 + (\gamma + \mu)i^*_2
\end{array} 
\right).
\label{B(4)} 
\end{equation}
Using Eq.~(\ref{two_cities__deter}) at the fixed point these may be 
simplified to
\begin{equation}
B^{(1)} = \left( 
\begin{array}{cc}
2\mu(1-s^*_1) & 0 \\ \\
0 & 2\mu(1-s^*_2)
\end{array} 
\right),
\label{B(1)_2} 
\end{equation}
\begin{equation}
B^{(2)} = B^{(3)} = \left( 
\begin{array}{cc}
-(\gamma+2\mu) i^*_1 & 0 \\ \\
0 & -(\gamma+2\mu) i^*_2 
\end{array} 
\right),
\label{B(2)B(3)_2} 
\end{equation}
and
\begin{equation}
B^{(4)} = \left( 
\begin{array}{cc}
2(\gamma + \mu)i^*_1  & 0 \\ \\
0 & 2(\gamma + \mu)i^*_2
\end{array} 
\right).
\label{B(4)_2} 
\end{equation}

\medskip

We now introduce the matrix
\begin{eqnarray}
P_{JK}(\omega) &\equiv& \langle \tilde{z}_J(\omega) 
\tilde{z}^*_K(\omega) \rangle \nonumber \\
&=& \sum^{4}_{L=1} \sum^{4}_{M=1} \Phi^{-1}_{JL}(\omega)B_{LM}
\left( \Phi^{\dag} \right)^{-1}_{MK}(\omega),
\label{defn}
\end{eqnarray}
where here (and only here) $*$ denotes complex conjugation. In the one-city 
case, where the focus is on finding the frequencies and amplitudes of the 
stochastic oscillations, only the power spectrum (when $J=K$) is usually 
analysed. When studying the model with two cities, we will also be interested 
in the cross-correlations between infection in two different cities, and so 
will also wish to calculate the cross-spectrum (when $J \neq K$). It is 
frequently convenient to normalise this by the relevant power-spectrum, and 
instead work with the complex coherence function (CCF) defined by
\begin{equation}
C_{JK}(\omega) \equiv \frac{P_{JK}(\omega)}{\sqrt{P_{JJ}(\omega)P_{KK}(\omega)}}.
\label{CCF}
\end{equation}
The CCF will in general be complex for $J \neq K$, and so typically one 
calculates its magnitude and phase. The phase is given by
\begin{equation}
\phi_{JK}(\omega) \equiv 
\tan^{-1}\left[ \frac{{\rm Im}\left( C_{JK}(\omega) \right)}
{{\rm Re}\left( C_{JK}(\omega) \right)} \right] =
\tan^{-1}\left[ \frac{{\rm Im}\left( P_{JK}(\omega) \right)}
{{\rm Re}\left( P_{JK}(\omega) \right)} \right].
\label{phase_spect}
\end{equation}
The phase used in the main text is found by evaluating (\ref{phase_spect}) at 
the value of $\omega$ that maximises the modulus of the CCF (\ref{CCF})~[4].

\medskip

\section{The $n$-City Stochastic SIR model}
We will be relatively brief in this section, and only outline the results, since
the formalism and general procedure is as for the two-city case. The main
difference is that there is a fifth type of infection event --- in addition to
those mentioned for two cities. This is due to the fact that infective 
individuals can commute from city $k$ and infect susceptible individuals from 
city $j$ in city $\ell$, where $j, k$ and $\ell$ are all different. This is 
only possible when there are three or more cities.

\medskip

\subsection{Form of the interaction between the $n$ cities}

The number of individuals in the three classes belonging to city $j$ are 
denoted by $S_{j}, I_{j}$ and $R_{j}=N_{j}-S_{j}-I_{j}$ as before, where now
$j=1,\ldots,n$. We will also introduce the notation
\begin{equation}
f_j = \sum_{k\neq j} f_{kj},
\label{f_j}
\end{equation}
so that the number of individuals in city $j$ may be written as
\begin{eqnarray}
M_j &=& \Bigl[ 1 - \sum_{k\neq j} f_{kj} \Bigr]N_j + \sum_{k\neq j} f_{jk}N_k 
\nonumber \\
&=& \left( 1 - f_j \right)N_j + \sum_{k\neq j} f_{jk}N_{k}.
\label{M_n}
\end{eqnarray}
As for the two city case, we will assume that the birth/death rate and the 
recovery rate are the same in both cities, but that the infection rate for
city $j$ is $\beta_j$.

As mentioned above, there are five different types of infection events:
\begin{itemize}
\item[(i)] Infective residents of one city (say $j$) can infect susceptible 
residents of the same city. The rate for this to occur is
$\beta_j\,\left( 1 - f_{j}\right)S_j\,\left( 1 - f_{j}\right)I_j/M_j$.
\item[(ii)] Infective commuters can travel from city $k$ to city $j$ and infect 
susceptible residents in their home city $j$. The rate for this to occur is
$\beta_j\,\left( 1 - f_{j}\right)S_j\,f_{jk}I_k/M_j$.
\item[(iii)] Infective residents in their home city (say $k$) can infect 
susceptible commuters from the other city ($j$). The rate for this to occur is
$\beta_k\,f_{kj}S_j\,\left( 1 - f_{k}\right)I_k/M_k$.
\item[(iv)] Infective commuters from city $j$ infect susceptible commuters 
from $j$ in city $\ell$ ($\ell \neq j$). The rate for this to occur is
$\beta_{\ell}\,f_{\ell j}S_j\,f_{\ell j}I_j/M_{\ell}$.
\item[(v)] Infective commuters from city $k$ can infect susceptible commuters
from city $j$ in city $\ell$ ($\ell \neq j,k$). The rate for this to occur is
$\beta_{\ell}\,f_{\ell j}S_j\,f_{\ell k}I_k/M_{\ell}$.
\end{itemize}

Adding these rates together we obtain the total transition rate for infection 
of $S_j$ individuals as 
\begin{equation}
\sum^{n}_{k=1} \beta_{jk} \frac{S_j I_k}{N_k},
\label{infection}
\end{equation}
where
\begin{eqnarray}
\beta_{jj} &=& \frac{\beta_j \left( 1 - f_j \right)^2 N_j}{M_j}
+ \sum_{\ell \neq j} \frac{\beta_{\ell}f_{\ell j}^2 N_j}{M_\ell}, \ \ j=1,\ldots,n,
\nonumber \\
\beta_{jk} &=& \frac{\beta_j \left( 1 - f_j \right)f_{jk} N_k}{M_j} 
+ \frac{\beta_k f_{kj}\left( 1 - f_k \right) N_k}{M_k} \nonumber \\
&+& \sum_{\ell\neq j,k} \frac{\beta_{\ell} f_{\ell j} f_{\ell k} N_k}{M_\ell}, \ \
j,k=1,\ldots,n; j \neq k.
\label{beta_jk_ncities}
\end{eqnarray}

\medskip

The analogue of Eq.~(\ref{t_twocity_rates}) is 
\begin{equation}
\begin{array}{l r l}
\textrm{1. Infection of $S_j$} & T_{1j}(S_{j}-1,I_{j}+1|S_j,I_j)\ = & 
\sum^{n}_{k=1} \beta_{jk} \frac{S_j I_k}{N_k} \\ \\
\textrm{2. Recovery of $I_j$} & T_{2j}(S_j,I_{j}-1|S_j,I_j)\ = & \gamma I_j \\ \\
\textrm{3. Birth/death in city $j$} & T_{3j}(S_{j}+1,I_{j}-1|S_j,I_j)\  
= & \mu I_j \\ \\
\textrm{4. Birth/death in city $j$} & T_{4j}(S_{j}+1,I_{j}|S_j,I_j)\  = & 
\mu (N_{j}-S_{j}-I_{j}),
\end{array}
\label{t_ncity_rates}
\end{equation}
although we have aggregated the reactions for convenience, so that each line
represents $n$ reactions with $j=1,\ldots,n$.

\medskip

From this, the information required to deduce the macroscopic and stochastic
equations can be found:
\begin{equation}
\begin{array}{l r l}
\textrm{1. Infection of $S_j$} & \nu^{(j)}_{1 j} = -1, \ \nu^{(n+j)}_{1 j} = +1 & 
g_{1 j} = \sum^{n}_{k=1} \beta_{jk} s_j i_k \\ 
\textrm{2. Recovery of $I_j$} & \nu^{(n+j)}_{2 j} = -1 & g_{2 j} = \gamma i_j \\
\textrm{3. Birth/death in city $j$} & \nu^{(j)}_{3 j} = +1, \ \nu^{(n+j)}_{3 j} 
= -1 & g_{3 j} = \mu i_j \\
\textrm{4. Birth/death in city $j$} & \nu^{(j)}_{4 j} = +1 & 
g_{4 j} = \mu(1 -s_{j}-i_{j}). 
\end{array}
\label{info_req_n_cities}
\end{equation}

\medskip

\subsection{The macroscopic and stochastic equations}
The macroscopic equation (\ref{macro_eqn}), where 
$\mathbf{A}(\boldsymbol{\phi})$ is defined by Eq.~(\ref{A_def}), when expressed 
in terms of the components of 
$\boldsymbol{\phi}=\left( s_1,\ldots,s_n,i_i,\ldots,i_n \right)$ may be 
written down using the information in Eq.~(\ref{info_req_n_cities}). One finds
\begin{eqnarray}
\frac{\mathrm{d}s_j}{\mathrm{d}t} &=& - \sum^{n}_{k=1} \beta_{jk} s_j i_k 
+ \mu\left( 1 - s_j \right), \ \ \ j=1,\ldots,n, \nonumber \\
\frac{\mathrm{d}i_j}{\mathrm{d}t} &=& \sum^{n}_{k=1} \beta_{jk} s_j i_k 
- \left(\gamma + \mu\right) i_j, \ \ \ j=1,\ldots,n.
\label{n_cities__deter}
\end{eqnarray}

\medskip

The Jacobian evaluated at the unique stable non-trivial fixed point is 
\begin{equation}
{\mathcal J} = \left[\begin{array}{c|c} 
{\mathcal J}^{(1)} & {\mathcal J}^{(2)} \\ \hline 
{\mathcal J}^{(3)} & {\mathcal J}^{(4)} 
 \end{array}\right],
\label{block_J}
\end{equation}
where these submatrices are given by
\begin{eqnarray}
{\mathcal J}^{(1)}_{jk} &=& - \left[ \sum^{n}_{\ell = 1} \beta_{j\ell} i^*_{\ell} 
+ \mu \right]\delta_{jk}, \ \ \ \ \ \ 
{\mathcal J}^{(2)}_{jk} = - \beta_{jk} s^*_j, \nonumber \\
{\mathcal J}^{(3)}_{jk} &=& \left[ \sum^{n}_{\ell = 1} 
\beta_{j\ell} i^*_{\ell} \right]\delta_{jk}, \ \ \ \ \ \ \ \ \ \ \ \ \ \ 
{\mathcal J}^{(4)}_{jk} = \beta_{jk} s^*_j - 
\left( \gamma + \mu\right)\delta_{jk}.
\label{Jacobian_n_cities}
\end{eqnarray}

\medskip

Finally, the $B$ matrix (\ref{B_def}) evaluated at a fixed point has the form
(\ref{block}) where
\begin{eqnarray}
B^{(1)}_{jk} &=& \left[ \sum^{n}_{\ell=1} \beta_{j\ell}s^*_{j}i^*_{\ell} 
+ \mu(1-s^*_j) \right]\delta_{jk}, \nonumber \\
B^{(2)}_{jk} &=& B^{(3)}_{jk} = \left[ - \sum^{n}_{\ell=1} 
\beta_{j\ell}s^*_{j}i^*_{\ell} - \mu i^*_j \right]\delta_{jk}, \nonumber \\
B^{(4)}_{jk} &=& \left[ \sum^{n}_{\ell=1} \beta_{j\ell}s^*_{j}i^*_{\ell} 
+ (\gamma + \mu)i^*_j \right]\delta_{jk}.
\label{B_n_cities}
\end{eqnarray}
Using Eq.~(\ref{n_cities__deter}) at the fixed point these may be 
simplified to
\begin{eqnarray}
B^{(1)}_{jk} &=& \left[ 2 \mu(1-s^*_j) \right]\delta_{jk}, \nonumber \\
B^{(2)}_{jk} &=& B^{(3)}_{jk} = - \left[ (\gamma + 2\mu) i^*_j \right]\delta_{jk},
 \nonumber \\
B^{(4)}_{jk} &=& \left[ 2 (\gamma + \mu)i^*_j \right]\delta_{jk}.
\label{B_simplified_n_cities}
\end{eqnarray}

\noindent The cross-spectra, complex coherence function and phase spectra are 
as in Eqs.~(\ref{defn})-(\ref{phase_spect}).

\medskip

\section{Additional results}
In this section, we will give additional results for the 1-city (Figure 1), 2-city (Figure 2) and 4-city (Figure 3) SIR models.

Figure 1 shows the average extinction time (AET) from simulations of the 1-city SIR model as a function of the basic reproductive ratio, $R_0$, for different population sizes, $N$, and levels of seasonality, $\epsilon$. The results for $N=4\times10^5$ (middle row) were shown in Figure 1 of the main text. We observe that for smaller ($N=2\times10^5$, top row)/larger ($N=8\times10^5$, bottom row) populations the dependence on $R_0$ of the AET becomes much less/more pronounced.

Figure 2 shows the AET from simulations of the 2-city SIR model for different levels of seasonality, $\epsilon$, and analytical results for the amplification, coherence for both cities and phase lag between cities. The basic reproductive ratio is taken to be $R_{01}=18+\delta$ in one of the cities and $R_{02}=18-\delta$ in the the other city, where $\delta=0,1,\ldots,6$. Panels (a), (b), (c), (e), (g) and (h) are repeated from Figure 3 of the main text. The new panels (d) and (f) show that both the coherence and amplification in the first city are low in the region where the AET is longest, similarly to the results for the second city, shown in panels (e) and (g).

Figure 3 shows the same results as Figure 1 but for the 4-city SIR model in which a central city with larger population is surrounded by 3 non-interacting satellite cities with smaller (and equal) populations. The basic reproductive ratio is taken to be $R_{01}=18+\delta$ in the larger city and $R_{02,03,04}=18-\delta$ in the smaller cities, where $\delta=0,1,\ldots,6$. The results refer to the central city and one of the satellite cities. We consider the case when the fraction of commuters from each of the satellite cities to the central city, $f$, is 10 times larger than that from the central city to that satellite city. Panels (a), (b), (c) repeated from Figure 5 of the main text show that the AET is maximum at strong heterogeneity (large $\delta$) and intermediate coupling strengths $f$. The additional panels (d)-(h) again suggest that this behaviour can be explained in terms of the coherence and amplification and that the phase lag between the cities does not play the dominant role in such behaviour of the AET.

\begin{figure*}
\centering
\includegraphics[trim=0cm 0cm 0cm 0cm, clip=true, width=0.361\textwidth]{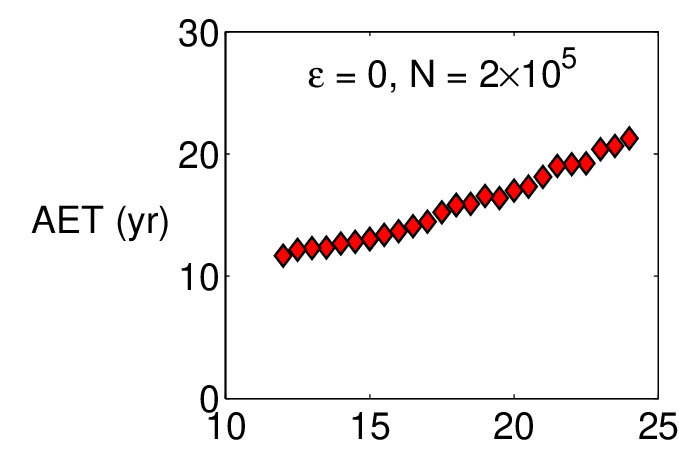}
\includegraphics[trim=0cm 0cm 0cm 0cm, clip=true, width=0.629\textwidth]{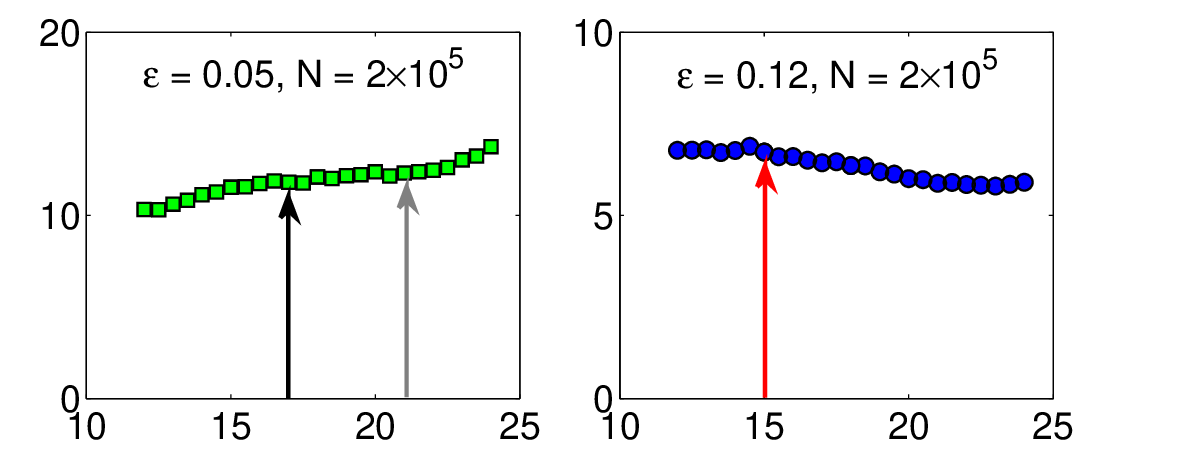}\\
\includegraphics[trim=0cm 0cm 0cm 0cm, clip=true, width=0.361\textwidth]{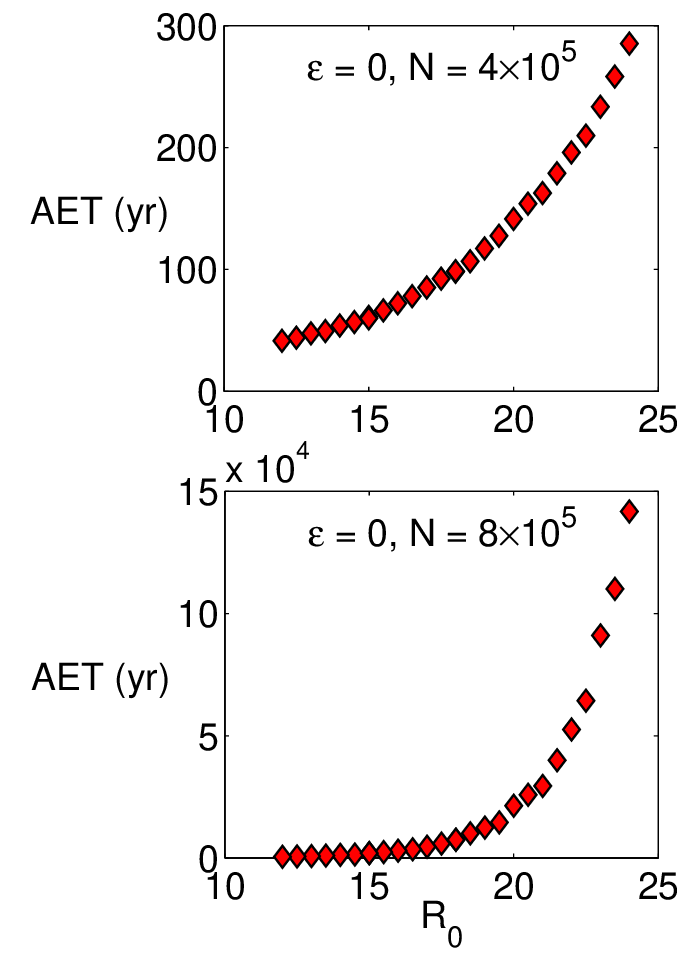}
\includegraphics[trim=0cm 0cm 0cm 0cm, clip=true, width=0.629\textwidth]{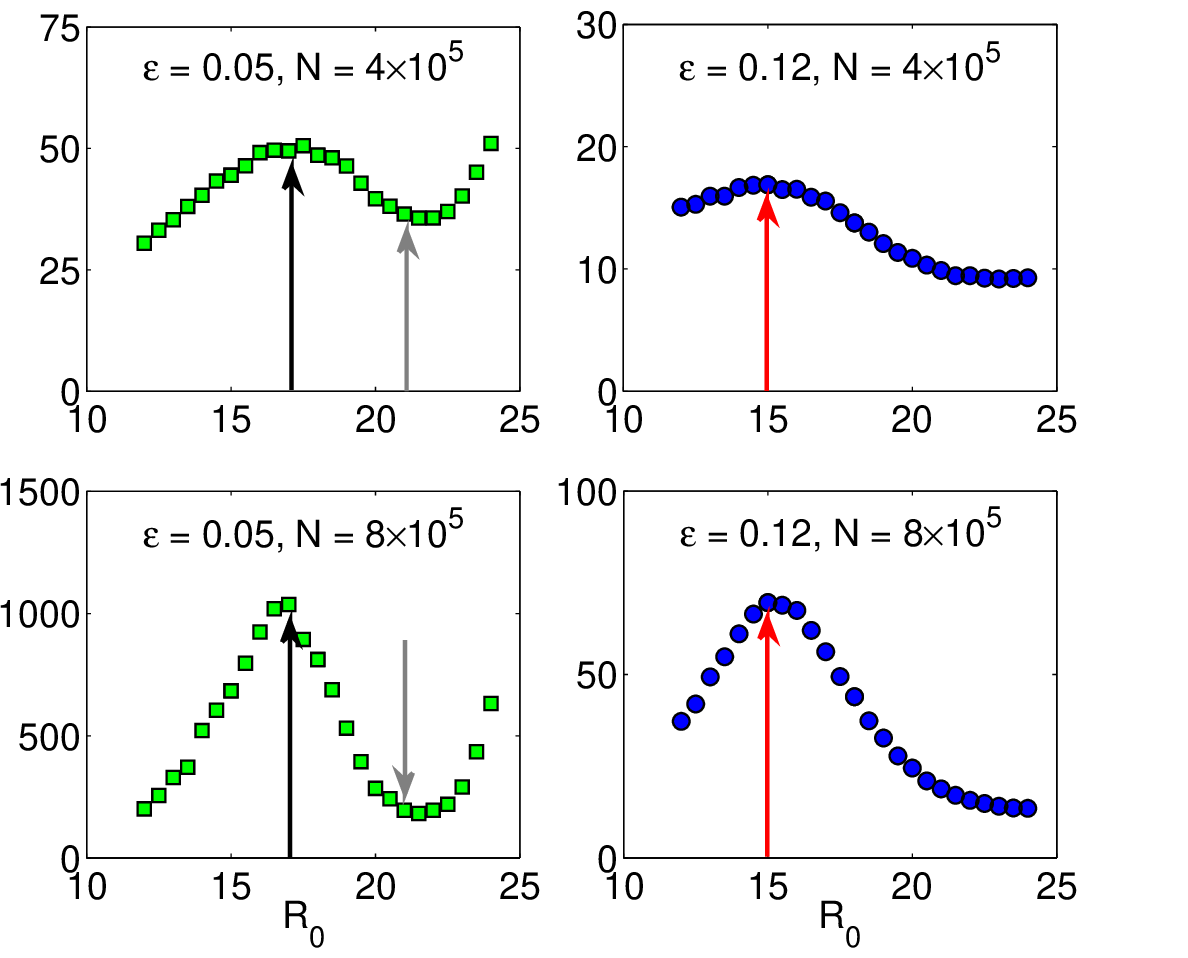}
{\caption{AET from simulations for the 1-city SIR model. The arrows highlight the parameter regions where the attractor of the forced system changes from annual to biennial (black and red arrows), and from biennial to annual (grey arrows).}}
\label{fig1}
\end{figure*}

\begin{figure*}
\centering
\includegraphics[trim=0cm 0cm 0cm 0cm, clip=true, width=0.403\textwidth]{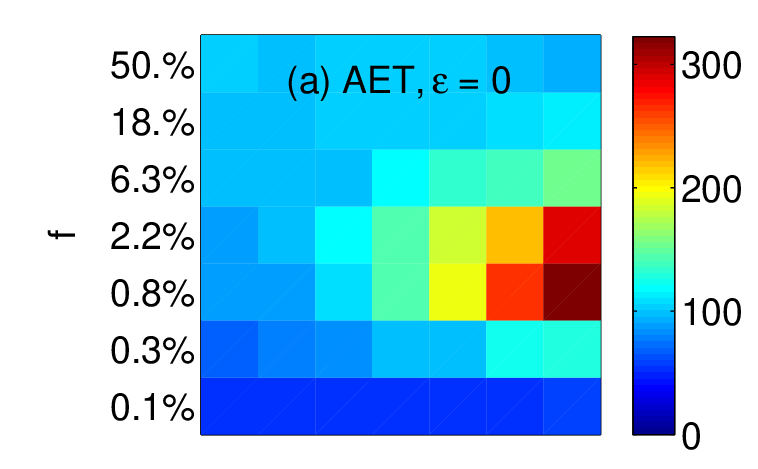}
\includegraphics[trim=0cm 0cm 0cm 0cm, clip=true, width=0.587\textwidth]{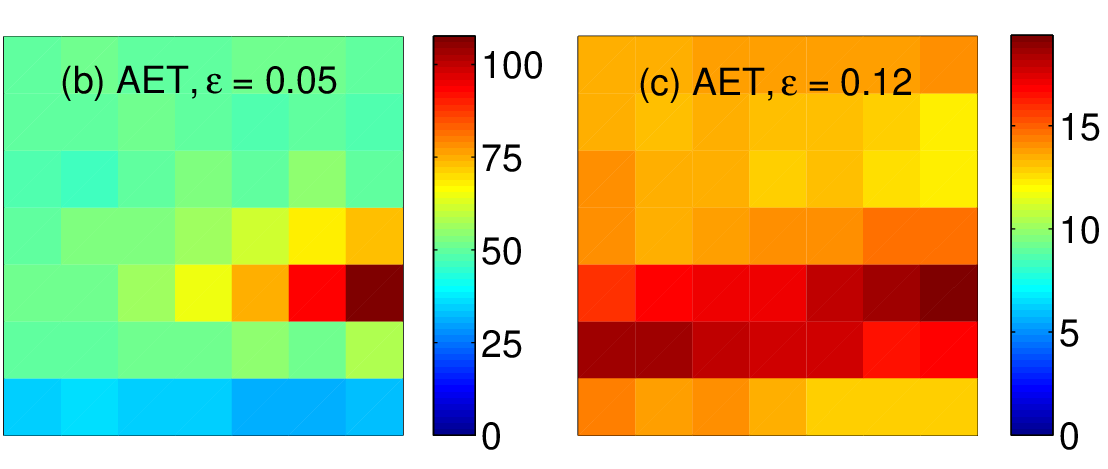}\\
\includegraphics[trim=0cm 0cm 0cm 0cm, clip=true, width=0.707\textwidth]{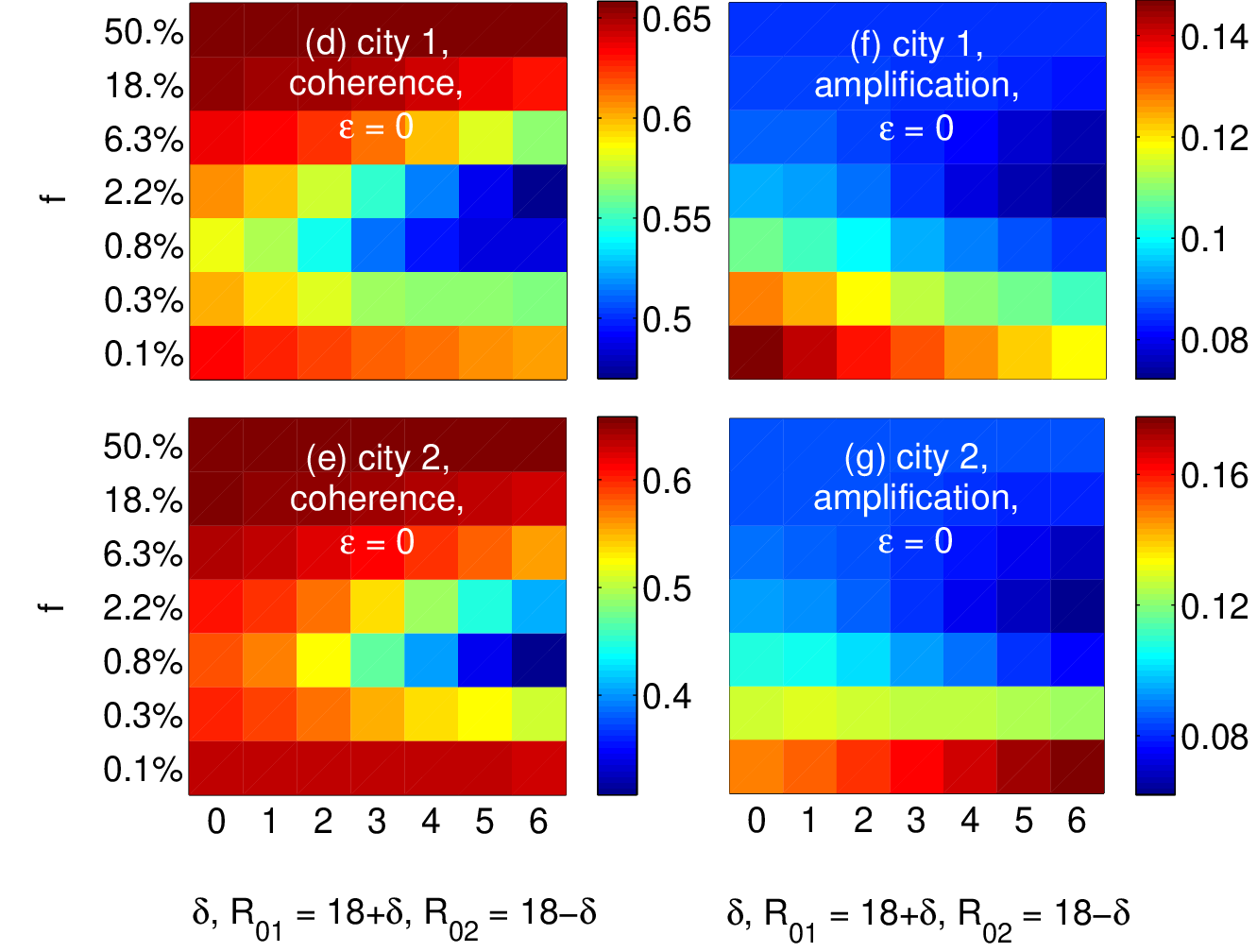}
\includegraphics[trim=0cm 0cm 0cm 0cm, clip=true, width=0.283\textwidth]{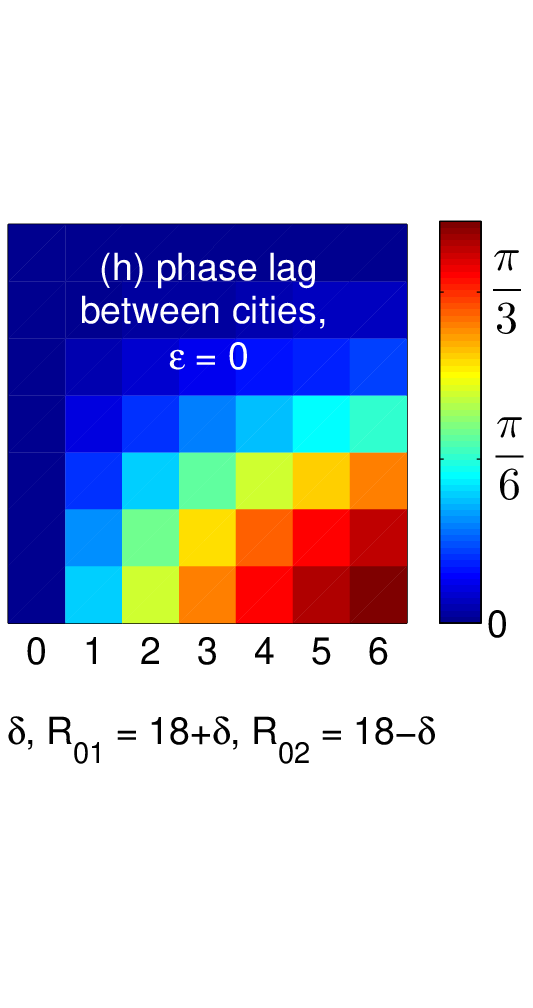}
{\caption{Results for the 2-city SIR model with $N_1=N_2=2\times10^5$. (a)-(c) AET from simulations for different levels of seasonality, $\epsilon$. Analytical results for (d)-(e) coherence, (f)-(g) amplification and (h) phase lag between cities for $\epsilon=0$.}}
\label{fig2}
\end{figure*}

\begin{figure*}
\centering
\includegraphics[trim=0cm 0cm 0cm 0cm, clip=true, width=0.403\textwidth]{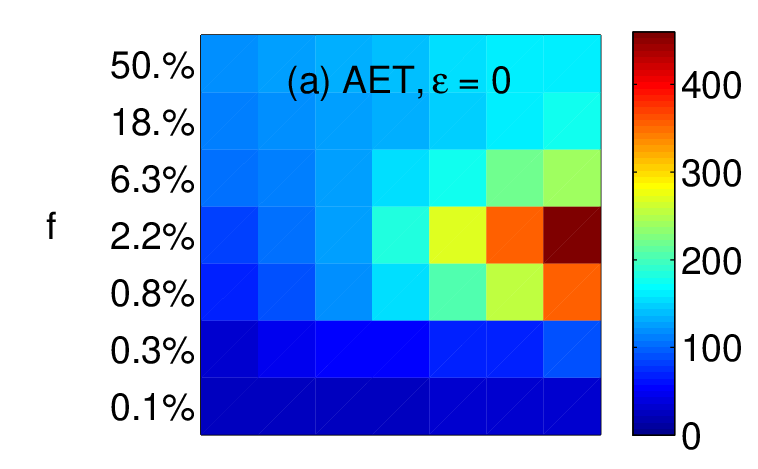}
\includegraphics[trim=0cm 0cm 0cm 0cm, clip=true, width=0.587\textwidth]{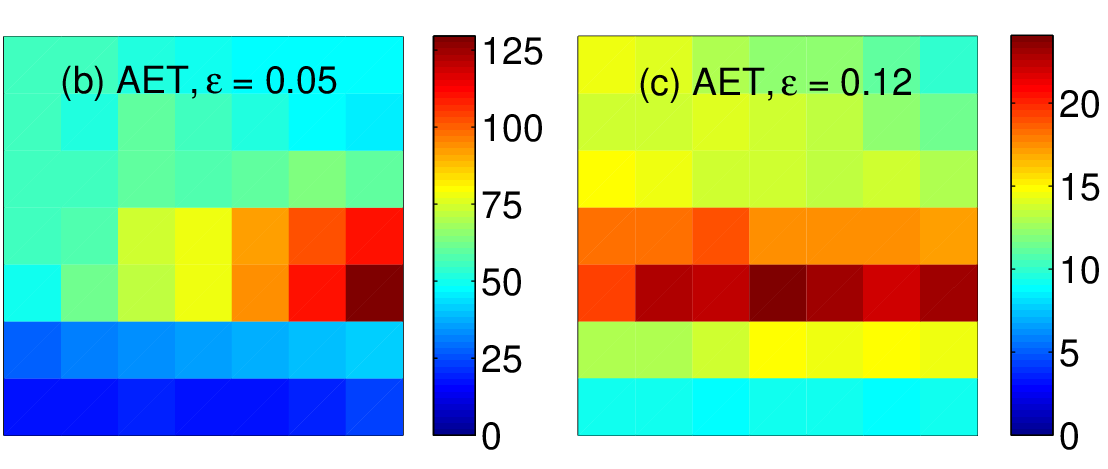}\\
\includegraphics[trim=0cm 0cm 0cm 0cm, clip=true, width=0.707\textwidth]{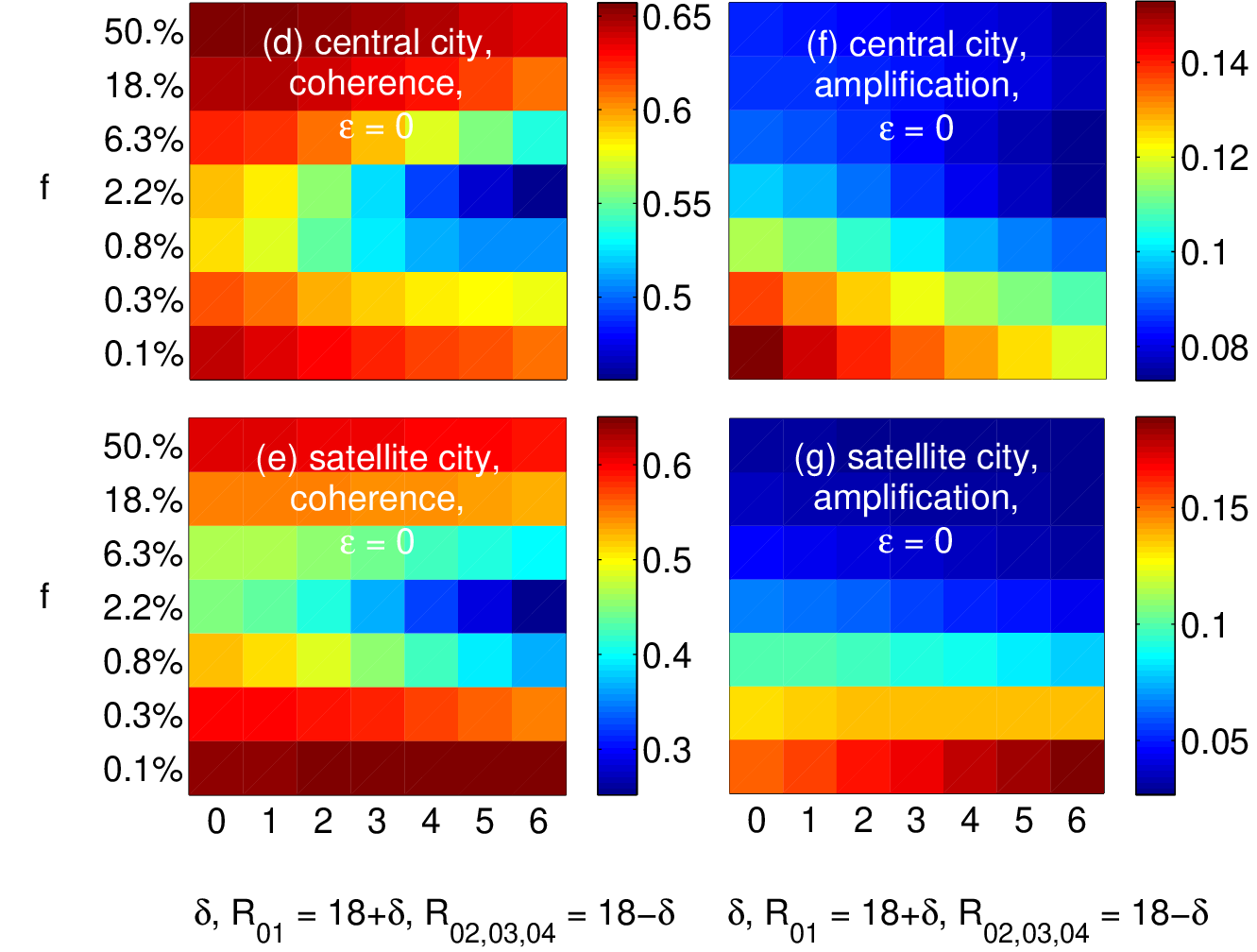}
\includegraphics[trim=0cm 0cm 0cm 0cm, clip=true, width=0.283\textwidth]{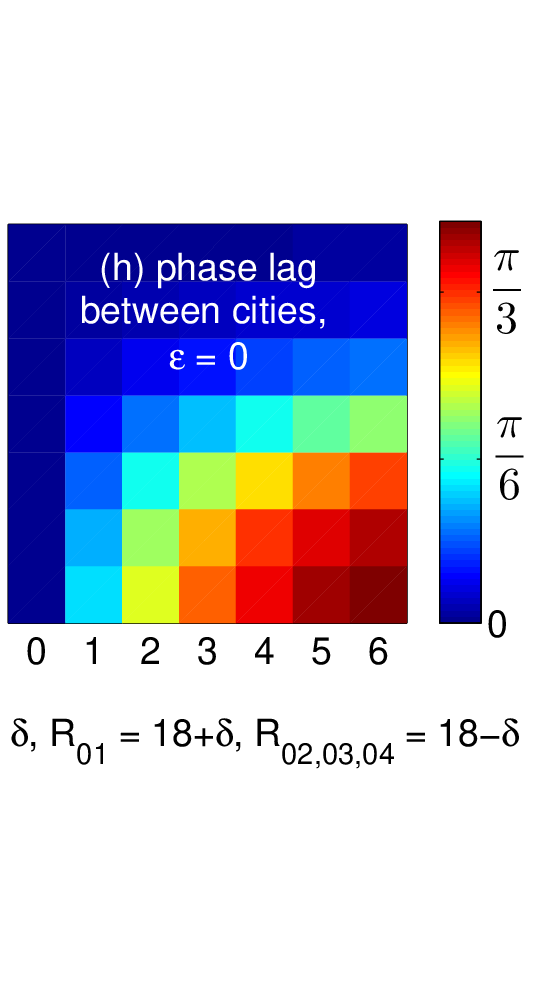}
{\caption{Results for the 4-city SIR model in which a central city with $N_1=2.1\times10^5$ is surrounded by 3 non-interacting satellite cities with $N_2=N_3=N_4=N_1/3=7\times10^4$. Fraction of commuters from each of the suburban cities to the central city, $f$, is 10 times larger than that from the central city to that suburban city. (a)-(c) AET from simulations for different levels of seasonality, $\epsilon$. Analytical results for (d)-(e) coherence, (f)-(g) amplification and (h) phase lag for $\epsilon=0$ between the central city and one of the satellite cities.}}
\label{fig3}
\end{figure*}\newpage

{\large \textbf{References}}

\medskip

\begin{enumerate}

\item D. Alonso, A. J. McKane and M. Pascual, ``Stochastic amplification
in epidemics'', J. R. Soc. Interface {\bf 4}, 575-582 (2007).

\item G. Rozhnova and A. Nunes, ``Stochastic effects in a seasonally forced 
epidemic model'', Phys. Rev. E {\bf 82}, 041906 (2010).

\item G. Rozhnova, A. Nunes, and A. J. McKane, ``Stochastic oscillations 
in models of epidemics on a network of cities'', Phys. Rev. E {\bf 84}, 
051919 (2011).

\item G. Rozhnova, A. Nunes, and A. J. McKane, ``Phase lag in epidemics on a 
network of cities'', Phys. Rev. E {\bf 85}, 051912 (2012).

\item A. J. Black and A. J. McKane, ``Stochastic formulation of ecological 
models and their applications'', Trends Ecol. Evol. {\bf 27}, 337-345 (2012).

\item A. J. McKane, T. Biancalani and T. Rogers, ``Stochastic pattern
formation and spontaneous polarisation: The linear noise approximation and 
beyond'', arXiv:1211.0462.

\end{enumerate}